\documentclass[aps,prb,secnumarabic, nobibnotes, twocolumn,superscriptaddress]{revtex4-1}

\usepackage{amsfonts}
\usepackage{mathrsfs}
\usepackage{amsmath}
\usepackage{color}
\usepackage{natbib}
\usepackage{graphicx}
\usepackage{bm}
\usepackage{amssymb}
\usepackage{xspace}
\usepackage{epstopdf}
\usepackage{dcolumn}
\usepackage{longtable}
\usepackage{multirow}
\usepackage[colorlinks=true, letterpaper=true, pdfstartview=FitV, linkcolor=black, citecolor=red, urlcolor=black]{hyperref}

\begin{document}

\title{Cat's cradle-like Dirac semimetals in  layer groups with multiple screw axes: Application to 2D Borophene and Borophane}

\author{Xiaotong Fan}
\affiliation {Beijing Key Laboratory of Nanophotonics and Ultrafine Optoelectronic Systems, School of Physics, Beijing Institute of Technology, Beijing 100081, China}

\author{Dashuai Ma}
\affiliation {Beijing Key Laboratory of Nanophotonics and Ultrafine Optoelectronic Systems, School of Physics, Beijing Institute of Technology, Beijing 100081, China}

\author{Botao Fu}
\email{fubotao2008@gmail.com}
\affiliation {Beijing Key Laboratory of Nanophotonics and Ultrafine Optoelectronic Systems, School of Physics, Beijing Institute of Technology, Beijing 100081, China}
\affiliation {College of Physics and Electronic Engineering, Center for Computational Sciences, Sichuan  Normal University, Chengdu, 610068, China}

\author{Cheng-Cheng Liu}
\email{ccliu@bit.edu.cn}
\affiliation {Beijing Key Laboratory of Nanophotonics and Ultrafine Optoelectronic Systems, School of Physics, Beijing Institute of Technology, Beijing 100081, China}

\author{Yugui Yao}
\affiliation {Beijing Key Laboratory of Nanophotonics and Ultrafine Optoelectronic Systems, School of Physics, Beijing Institute of Technology, Beijing 100081, China}

\begin{abstract}
Recently, the crystal symmetry-protected topological semimetals have aroused extensive interests, especially for the nonsymmorphic symmetry-protected one. We list the possible nonmagnetic topological semimetals and develop their $k{\cdot}p$ Hamiltonian in all layer groups with multiple screw axes in the absence of spin-orbital coupling. We find a novel cat's cradle-like topological semimetal phase, which looks like multiple hourglass-like band structures staggered together. Furthermore, we propose the monolayer borophene and borophane with $pmmn$ layer group as the first material class to realize such novel semimetal phase. A pair of tilted anisotropic Dirac cones at the Fermi level is revealed in the two-dimensional boron-based materials and the low-energy effective models are given. Moreover, akin to three-dimensional Weyl semimetal, the topological property of these cat's cradle-like Dirac semimetals can be verified by the calculation of quantized Berry phase and the demonstration of flat Fermi-arc edge state connecting two Dirac points as well. Our finding that borophene and borophane are cat's cradle-like Dirac semimetals is of great interest for experiment and possible applications in the future.
\end{abstract}
\maketitle

\section{Introduction}
The existence of massless Dirac fermion in graphene has brought about novel properties\cite{RMPgra,1,2,3,5} such as ultrahigh carrier mobility, half-integer quantum Hall effects, Klein tunneling, etc.
The concept of semimetal state has been extended from Dirac semimetal to nodal line semimetal\cite{DNL1,DNL2}, which has a loop of gapless crossing points in the Brillouin zone (BZ).
It is known that the crystal symmetry plays a crucial role for the emergence of various topological semimetals in both two-dimensional (2D) and three-dimensional (3D) systems\cite{22,fang2012multi,wang2016hourglass,po2017symmetry}.
For example, the space inversion and time reversal symmetries protect Dirac cones in 2D systems\cite{8} and Dirac nodal lines in 3D systems\cite{PDNL1,PDNL2,17}, the mirror symmetry protect Dirac nodal lines on the mirror-invariant plane\cite{cusi,19,20}.

Recently, nonsymmorphic symmetries, including slide mirror or screw axes operations, are discovered to guarantee the so-called symmetry-enforced semimetal with certain integer electron filling, which is unavoidably formed because of the unique band connectivity\cite{21,wangAFM2017,wiedercat2016}.
For instance, a screw axis can protect a two-dimensional four-fold-degenerate Dirac point at the BZ boundary in the presence of spin-orbit coupling (SOC)\cite{22}, which has been proposed in monolayer HfGeTe\cite{23}.
On the other hand, another kind of nonsymmorphic semimetal that originates from band inversion and is protected by one screw axis has been proposed and realized in monolayer WTe$_2$\cite{nonsm2016} in the absence of SOC.
Motivated by those remarkable works, we are going to ask whether there are new nonsymmorphic semimetals when two or more screw axes are presented.

In this paper, we generalize the nonsymmorphic semimetals due to band inversion mechanism from one screw axis to two or more screw axes for centrosymmetric and non-centrosymmetric systems, respectively. Both hourglass-like and cat's cradle-like Dirac semimetals, as well as nodal line semimetals are obtained.
Those topological semimetal phases are clearly exemplified in 2D borophene (B$_8$)\cite{zhouxfB8} and borophane (B$_{2}$H$_{2}$)\cite{jiaoB2H2,b2h2xu,prb2018B2H2} with $pmmn$ layer group. The emergence of different semimetal phases can be indicated by the eigenvalues of screw and inversion operations for the inverted bands and the corresponding topological properties are clearly verified by the quantized $\pi$ Berry phase in the bulk states and flat Fermi-arc edge states connecting two Dirac points.
The $Z_2$ number in the absence of SOC can predict the topological property when the SOC effect is included and the nontrivial gap is opened at the Dirac point. We believe our results may stimulate further research interest on 2D Dirac materials with nonsymmorphic symmetries.

\section{nonsymmorphic semimetals from band inversion}
There are twelve layer groups with equal to or more than two screw axes in all the 80 layer groups. They are $p2_12_12$ (No. 21), $c222$ (No. 22), $pbam$ (No. 44), $pbma$ (No. 45), $pmmn$ (No. 46), $cmmm$ (No. 47), $cmme$ (No. 48), $p42_12$ (No. 54), $p\bar{4}2_1m$ (No. 58), $p\bar{4}b2$ (No. 60), $p4/mbm$ (No. 63), and $p4/nmm$ (No. 64), among which three layer groups (Nos. 54, 63, and 64) own four screw axes. However, in such three layer groups two of the four screw axes are along the diagonal direction, which does not result in the double degeneracy on the BZ boundary discussed below. Therefore, the number of the effective screw axes is still two.  In fact, the two screw axes is also along the diagonal direction in the layer group (No. 60), which is thus ignored below. Depending on the presence or absence of inversion symmetry, the left eleven layer groups fall into two categories. The category I includes the layer groups (Nos. 44-48, 63, 64) with inversion symmetry, while the category II includes Nos. 21, 22, 54, 58.  Without loss of generality, we take the simplest
representative cases from each of the two categories, i.e.,  $p$2$_1$2$_1$2 (No. 21) and $pmmn$ (No. 46), which only include two generators ${{\widetilde{C}}_{2x}}$ and ${{\widetilde{C}}_{2y}}$, and three generators ${{\widetilde{C}}_{2x}}$, ${{\widetilde{C}}_{2y}}$ and $P$, respectively.

Here we first focus on the centrosymmetric layer group $pmmn$. The generators of $pmmn$ contain two screw operations  (${{\widetilde{C}}_{2x}}=\{{{C}_{2x}}|a/2\}$, ${{\widetilde{C}}_{2y}}=\{{{C}_{2y}}|b/2\}$) and
an inversion operation $P$.
Other symmetry operations of $pmmn$ can been created as follow:
\begin{eqnarray}
\widetilde{C}_{2z}=\widetilde{C}_{2x}\widetilde{C}_{2y}, \widetilde{M}_{z}={P}(\widetilde{C}_{2x}\widetilde{C}_{2y}), \nonumber  \\
\widetilde{M}_{x}={P}\widetilde{C}_{2x}, \widetilde{M}_{y}={P}\widetilde{C}_{2y}. \label{Relat1}
\end{eqnarray}
Taking advantages of two screws axes and time reversal symmetry $T$,
we first demonstrate the symmetry-enforced double-degeneracy of Bloch states along the BZ boundaries which in fact forms a nodal line.
Then we consider band inversion around $\Gamma$ point and acquire an hourglass-like and cat's cradle-like Dirac semimetal states.

\begin{figure}
\includegraphics[width=8cm]{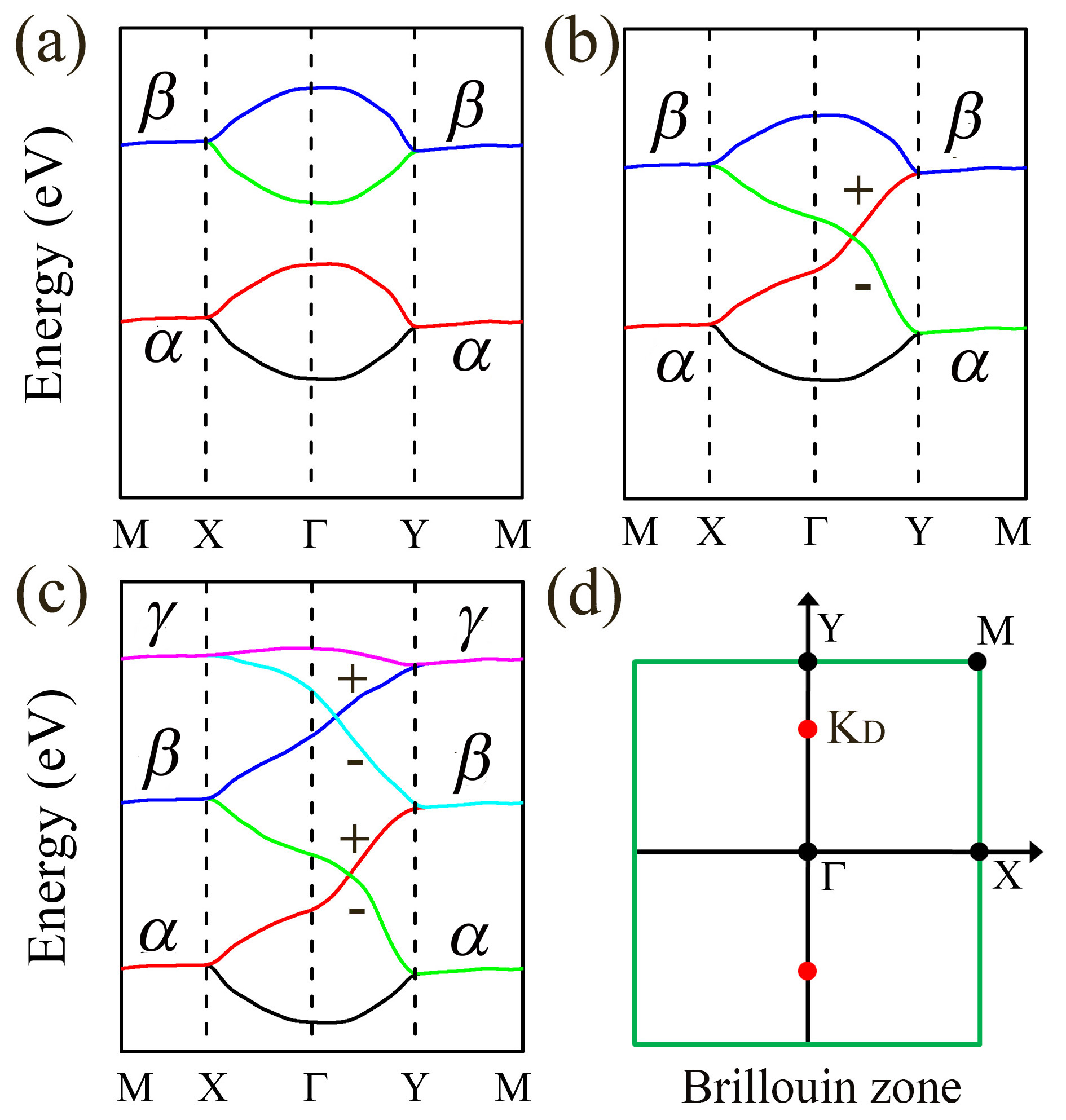}
\caption{Unique band connections for the nonsymmorphic layer groups with two screw axes, such as $p$2$_1$2$_1$2 and $pmmn$. (a) No band inversion happens. (b) Band inversion happens between pairs $\alpha$ and $\beta$ to form an hourglass-like band structure. (c) Band inversions happen simultaneously between pairs $\alpha$ and $\beta$, $\beta$ and $\gamma$, which gives a cat's cradle-like band structure. The positive or negative sign indicates the eigenvalue of screw operation along $\Gamma$Y path. (d) The first Brillouin zone. The green line represents double-degenerate $k$-point. The red point stands for location of Dirac point originated from band inversion mechanism at the Fermi level, labeled by $K_D$.
}\label{band1}
\end{figure}

Defining joint symmetry operations $\widetilde{T}_{x}$=$\widetilde{C}_{2x}T$ and $\widetilde{T}_{y}$=$\widetilde{C}_{2y}T$ respectively, and taking $\widetilde{T}_{x}$ as example, it has the relation:
\begin {equation}
{{\widetilde{T}}_{x}}H({{k}_{x}},{{k}_{y}})\widetilde{T}_{x}^{-1}=H(-{{k}_{x}},{{k}_{y}}) .
\end{equation}
When $k_x=0$ or $\pi$, $\widetilde{T}_{x}$ communicates with $H(k)$.
On the other hand, we have $\widetilde{T}_{x}^{2}={{e}^{i{{k}_{x}}}}$. Thus, along MX path ($k_x=\pi$), $\widetilde{T}_{x}^{2}=-1$ will result in Kramers-like degeneracy. Similarly, $\widetilde{T}_{y}$ can protect the two-fold degeneracy along MY path.
Therefore, as shown in  Fig.~\ref{band1}, all bands are double degenerate forming nodal lines at BZ boundaries but are splitting inside the BZ.
For a system with electron number of $2n$ (ignore the spin degrees of freedom), if there is no band inversion happening between two pairs of bands, the system is insulating as shown Fig.~\ref{band1}(a). However, if band inversion happens at $\Gamma$ point between two pairs (labelled as $\alpha$ and $\beta$) of bands, a pair of Dirac points protected by the screw axis will emerge as shown in Fig.~\ref{band1}(b). The band structure shows an interesting hourglass-like dispersion, and we note it as an hourglass-like Dirac semimetal when the Fermi level crosses the Dirac point.
Moreover, if band inversions happen between three pairs of bands (labelled as $\alpha$, $\beta$ and $\gamma$), two hourglass-like structures will stack up and form a cat's cradle-like Dirac semimetal as shown in Fig.~\ref{band1}(c).
It is worth noticing that those two topological semimetal states are only protected by two screw axes and time reversal symmetry, irrelevant with the inversion symmetry. Thus, such novel topological semimetal states can also be found for 2D materials with layer groups including  equal to or more than two screw axes.

\section{Classification of nonsymmorphic semimetals}

Based on the eigenvalues of three independent operations ($\widetilde{C}_{2x}$, $\widetilde{C}_{2y}$, ${P}$), we are going to give all possible symmetry-protected semimetal phases for the $pmmn$ layer group.
The eigenvalues of a symmetry operation for two inverted bands can either be the same or opposite.
With three operations, we can get eight different cases in principle.
By discarding  two equivalent cases because of equivalence of $x$ and $y$ directions and one case that
can not support semimetal state, we actually discover five distinctive topological semimetal states listed in Table~\ref{Table.str}. The eigenvalues of $\widetilde{C}_{2x}$, $\widetilde{C}_{2y}$, ${P}$ are given and the eigenvalues of $\widetilde{M}_{x}$, $\widetilde{M}_{y}$ and $\widetilde{M}_{z}$ are derived from Eq. ~(\ref{Relat1}). The $Z_2$ numbers can be calculated based on the famous parity criterion\cite{FuLparity2007}, where $Z_2=1$ means it is a topological insulator and $Z_2=0$ means it is a normal insulator when SOC effect is taken into consideration and a gap is inevitably opened.

\begin{table}[t]
\caption{Five distinctive topological semimetal states are listed. For each operation, its eigenvalues for two inverted bands at $\Gamma$ point are given.
 The situations of band crossings and $Z_2$ numbers are shown in the last column.}
\vspace{0.2cm}
\renewcommand\arraystretch{1.6}
\begin{tabular}{p{0.6cm}<{ \centering}p{0.85cm}<{\centering}p{0.85cm}<{\centering}p{0.85cm}<{\centering}p{0.85cm}<{\centering}p{0.85cm}<{\centering}p{0.85cm}<{\centering}p{2cm}}
\hline
\hline
 $types  $ & $\widetilde{C}_{2x}$ & $\widetilde{C}_{2y}$ & ${P}$ & $\widetilde{M_{x}} $ & $\widetilde{M_{y}} $ & $\widetilde{M}_{z}$ &    $result$     \\
\hline
$\mathrm{I}$ & $+,+$& $ +,- $&$ +,- $& $+,- $  & $+,+ $ & $+,+ $ &  $  2 DPs,Z_{2}=1 $ \\
\hline
$\mathrm{II}$ & $ +,- $&$ +,- $&   $+,+$& $ +,- $&$ +,- $& $+,+ $& $  4 DPs,Z_{2}=0  $  \\
\hline
$\mathrm{III}$ & $+,+$& $ +,+ $&$ +,- $& $+,- $  & $+,- $ & $+,- $ & $  DNL,Z_{2}=1  $   \\
\hline
$\mathrm{IV}$ & $+,-$& $ +,- $&$ +,- $& $+,+ $  & $+,+ $ & $+,- $ & $  DNL,Z_{2}=1  $  \\
\hline
$\mathrm{V}  $ & $ +,- $  & $+,+ $ & $+,+$ & $+,- $  & $+,+ $   & $+,-$ & $  DNL,Z_{2}=0  $    \\
\hline

\end{tabular}\label{Table.str}
\end{table}

Moreover, we are going to give two-band $k{\cdot}p$ models centered on $\Gamma$ point for the five topological semimetal states. Generally, an effective $k{\cdot}p$ Hamiltonian can be expressed as:
\begin{eqnarray}
H\left(\boldsymbol{k}\right)=f_0\left(\boldsymbol{k}\right)\tau_{0}+
f_1\left(\boldsymbol{k}\right)\tau_{x}+
f_2\left(\boldsymbol{k}\right)\tau_{y}+
f_3\left(\boldsymbol{k}\right)\tau_{z}, \label{Halmit0}
\end{eqnarray}
where $\tau_0$ is the identity matrix and $\tau_{x,y,z}$ are the Pauli matrices.
The time reversal operation can be taken as a complex conjugation operator for the SU(2) symmetry, and a crystal symmetry operation can be written as $\tau_z$ ($\tau_0$) if its eigenvalues have the opposite (same) signs for two bands at $\Gamma$ point.
The constrains from those symmetries require the following conditions:
\begin{eqnarray}
{T}H\left(\boldsymbol{k}\right){T}^{-1}=H\left(-\boldsymbol{k}\right),\nonumber  \\
{P}H\left(\boldsymbol{k}\right){P}^{-1}=H\left(-\boldsymbol{k}\right),\nonumber  \\
\widetilde{C}_{2x}H\left(k_{x},k_{y}\right)\widetilde{C}_{2x}^{-1}=H\left(k_{x},-k_{y}\right),\nonumber   \\
\widetilde{C}_{2y}H\left(k_{x},k_{y}\right)\widetilde{C}_{2y}^{-1}=H\left(-k_{x},k_{y}\right).\label{Const1}
\end{eqnarray}

Combing Eqs.~(\ref{Halmit0}) and ~(\ref{Const1}), we can finally obtain the $k{\cdot}p$ models for the five topological semimetal phases in Table~\ref{Table.str}.

\begin{figure}
\includegraphics[width=3.5 in]{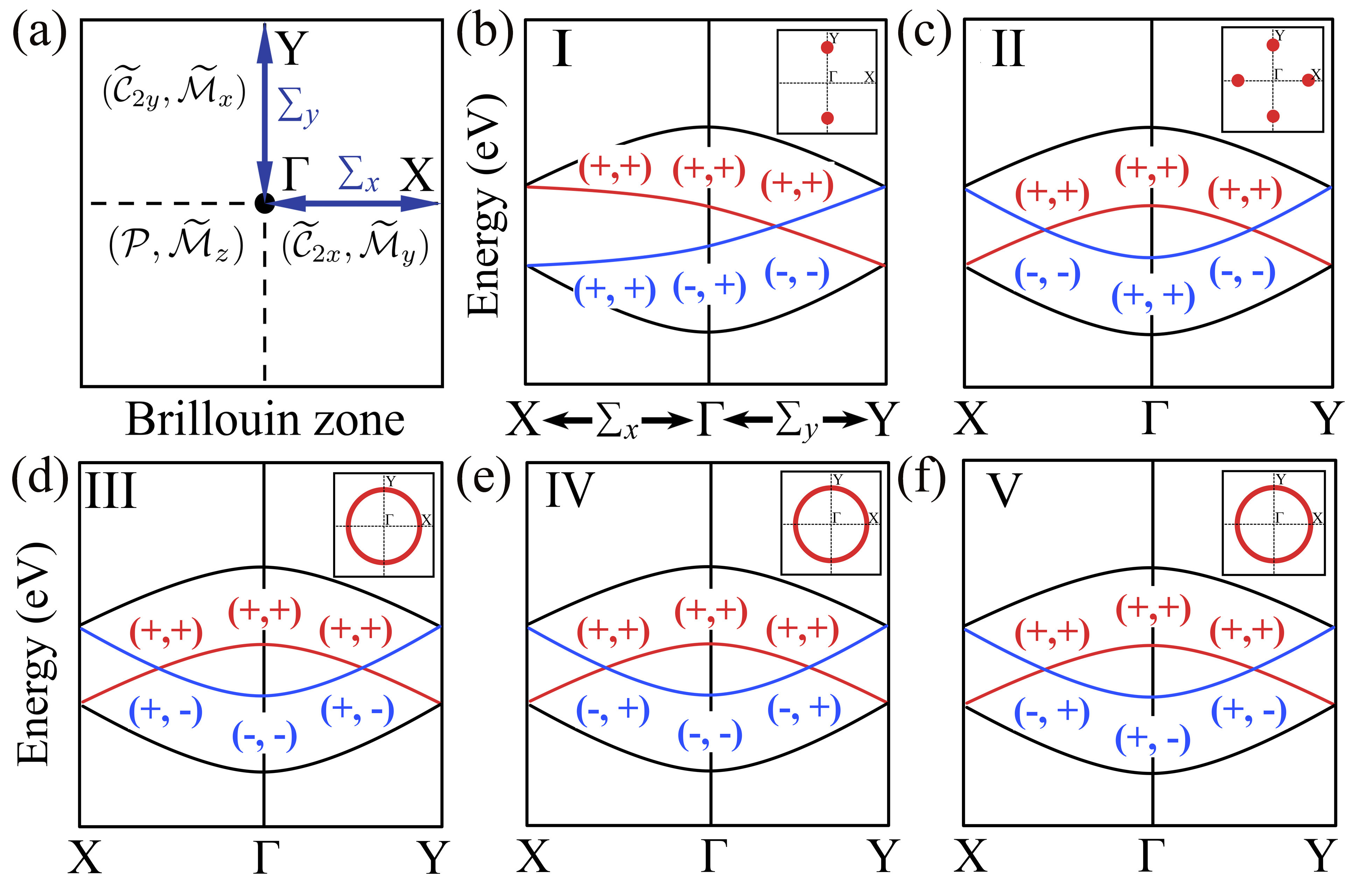}
\caption{(color online). (a) The first BZ with high symmetry points and lines. $\Sigma_y$ are invariant under $\widetilde{C}_{2y}$ and $\widetilde{M}_{x}$,  $\Sigma_x$ is invariant under $\widetilde{C}_{2x}$ and $\widetilde{M}_{y}$. (b)-(f) The schematic bandstructures for five distinctive topological semimetal states of $pmmn$ layer group, respectively.
In the brackets, the eigenvalues of corresponding operations displayed in (a) are given for the inverted bands. Each inset shows the band crossing points inside the BZ.
}\label{band2}
\end{figure}

The band structures for five topological semimetal states are exhibited in Fig.~\ref{band2}. Specifically, for type-I case in Fig.~\ref{band2}(b), two bands are inverted at $\Gamma$ point that forms an hourglass-like Dirac semimetal as predicted in Fig.~\ref{band1}(b).
Noting that the cat's cradle-like Dirac semimetal state shown in Fig.~\ref{band1}(c) also belongs to type-I case when similar band inversion processes happen between three pairs of bands.
A pair of Dirac points located along $\Gamma$Y path are individually protected by $\widetilde{C}_{2y}$ and $\widetilde{M}_{x}$. The Dirac points are robust as long as any one of those symmetries is reserved. For instance, imposing a vertical electric field breaks $\widetilde{C}_{2y}$ symmetry but leaves $\widetilde{M}_{x}$ symmetry, and the Dirac points still survive. The $Z_2=1$ for type-I case in the Table I indicates it will become a topological insulator when the SOC effect is included, which breaks SU(2) symmetry and opens a topological nontrivial gap.
The $k{\cdot}p$ model Hamiltonian approximated to the second order for type-I case is written as:
\begin{eqnarray}
H_{I}=f_{0}\tau_{0}+t_{1}k_{x}\tau_{y}+(m_{1}k_{x}^{2}+m_{2}k_{y}^{2}-m_{0})\tau_{z}, \label{Hkp1}
\end{eqnarray}
where the $t_i$, $m_i$ and $n_i$ in Eqs.(5)-(7) are fitting parameters.
The first term $f_{0}\tau_{0}$ would not change the topological property of Hamiltonian but can modify the shape of bandstructure, i.e., tilting effect.
It is obvious to see that above Hamiltonian will give a pair of Dirac cones at ($k_{x0}=0$, $k_{y0}=\pm \sqrt{m_{0}/m_{2}}$).

In Fig.~\ref{band2}(c), for type-II case, there are two pairs of Dirac points located along $\Gamma$X and $\Gamma$Y paths, respectively.
The Dirac point along $\Gamma$Y is individually protected by $\widetilde{C}_{2y}$ and $\widetilde{M}_{x}$, while the Dirac point along $\Gamma$X is individually protected by the $\widetilde{C}_{2x}$ and $\widetilde{M}_{y}$.
Similarly, a vertical electric field can not destroy those Dirac points because of remaining $\widetilde{M}_{x}$ and $\widetilde{M}_{y}$ symmetries.

We can see its $Z_2=0$ in the Table I, because two inverted bands have the same parities at $\Gamma$ point.
Thus the system will be a normal insulator when considering SOC effect.
The $k{\cdot}p$ model Hamiltonian for type-II case can be written as:
\begin{eqnarray}
H_{II}=f_{0}\tau_{0}+n_{1}k_{x}k_{y}\tau_{x}+(m_{1}k_{x}^{2}+m_{2}k_{y}^{2}-m_{0})\tau_{z}, \label{Hkp2}
\end{eqnarray}
which gives one pair of Dirac cones at ($k_{x0}=\pm \sqrt{m_{0}/m_{1}}$, $k_{y0}=0$) and  the other pair of Dirac cones at ($k_{x0}=0$, $k_{y0}=\pm \sqrt{m_{0}/m_{2}}$).

For type-III, IV, V cases, a Dirac Nodal line (DNL) state emerges as shown in Figs.~\ref{band2}(d),(e),(f).
The DNL is protected by the opposite eigenvalues of $\widetilde{M}_{z}$ operation.
The corresponding $k{\cdot}p$ model Hamiltonian for those three types can be written as:
\begin{eqnarray}
H_{III,IV,V}=f_{0}\tau_{0}+(m_{1}k_{x}^{2}+m_{2}k_{y}^{2}-m_{0})\tau_{z},  \label{Hkp3}
\end{eqnarray}
which can give a loop of band crossing points centered at $\Gamma$ point, namely a nodal line, in the BZ.

Although, those three DNL phases share the same $k{\cdot}p$ model, they response differently to external perturbations. For example, imposing a vertical electric field will break $\widetilde{M}_{z}$, ${P}$, $\widetilde{C}_{2x}$ and $\widetilde{C}_{2y}$, but leave $\widetilde{M}_{x}$ and $\widetilde{M}_{y}$.
As a consequence, a general mass term $\delta H$ is introduced and the DNL structure is destroyed.
For type-III case, the mass term $\delta H_{III}=\Delta_{3}k_{x}k_{y}\tau_{x}$ is introduced to Eq.~(\ref{Hkp3}),
and the new Hamiltonian $H_{III}+\delta H_{III}$ gives two pairs of Dirac points, which are protected by the opposite eigenvalues of $\widetilde{M}_{x}$ and $\widetilde{M}_{y}$ operations respectively.
For type-V case, the mass term $\delta H_{V}=\Delta_{5}k_{x}\tau_{y}$ is introduced to Eq.~(\ref{Hkp3}),
and the new Hamiltonian $H_{V}+\delta H_{V}$ presents a pair of Dirac points because of opposite eigenvalues of $\widetilde{M}_{x}$ operations.
Finally, for type-IV case, an additional term $\delta H_{IV}=\Delta_{4}\tau_{x}$ is introduced to Eq.~(\ref{Hkp3}), and $H_{V}+\delta H_{V}$ depicts a normal insulating state.

For the non-centrosymmetric layer groups $p$2$_1$2$_1$2 in category II, two screw axes $\widetilde{C}_{2x}$ and $\widetilde{C}_{2y}$ can only protect Dirac points along $\Gamma$X and $\Gamma$Y paths, respectively when the band inversion happens. Based on the eigenvalues of those two operations, it also gives type-I and type-II semimetal phases as listed in Table~\ref{Table.str}.

\section{Application to Borophene and Borophane}
Recently, an atomically thin crystalline, 2D boron sheets named as borophene with different forms \cite{borophenesy,fengsy,40,zhang2016sy,zhong2017sy,prmribbbionsy,wukehui}, have been experimentally synthesized. Especially, a Dirac cone was observed in boron sheet according to ARPES measurements\cite{fengprl,feng2018discovery}.
Meanwhile, several 2D Dirac materials with Dirac cones have been proposed in boron-based materials\cite{jiaoB2H2,b2h2xu,prb2018B2H2,prbborane,ma2016graphene,zhang2016dirac,TiB2dirac}.
Among them, the 2D borophene (B$_8$) and borophane(B$_2$H$_2$) sharing the nonsymmorphic layer group of $pmmn$ which exactly contains two screw axes in $x$ and $y$ directions.
In the following, we applying our results to B$_8$ and B$_{2}$H$_{2}$ systems and discover the cat's cradle-like and hourglass-like semimetal states in those two materials.

\renewcommand\thesection{\Roman{section}} \renewcommand\thesubsection{\Alph{subsection}}

\subsection{Crystal structures of B$_8$ and B$_{2}$H$_{2}$}
\begin{figure}
\includegraphics[width=8cm]{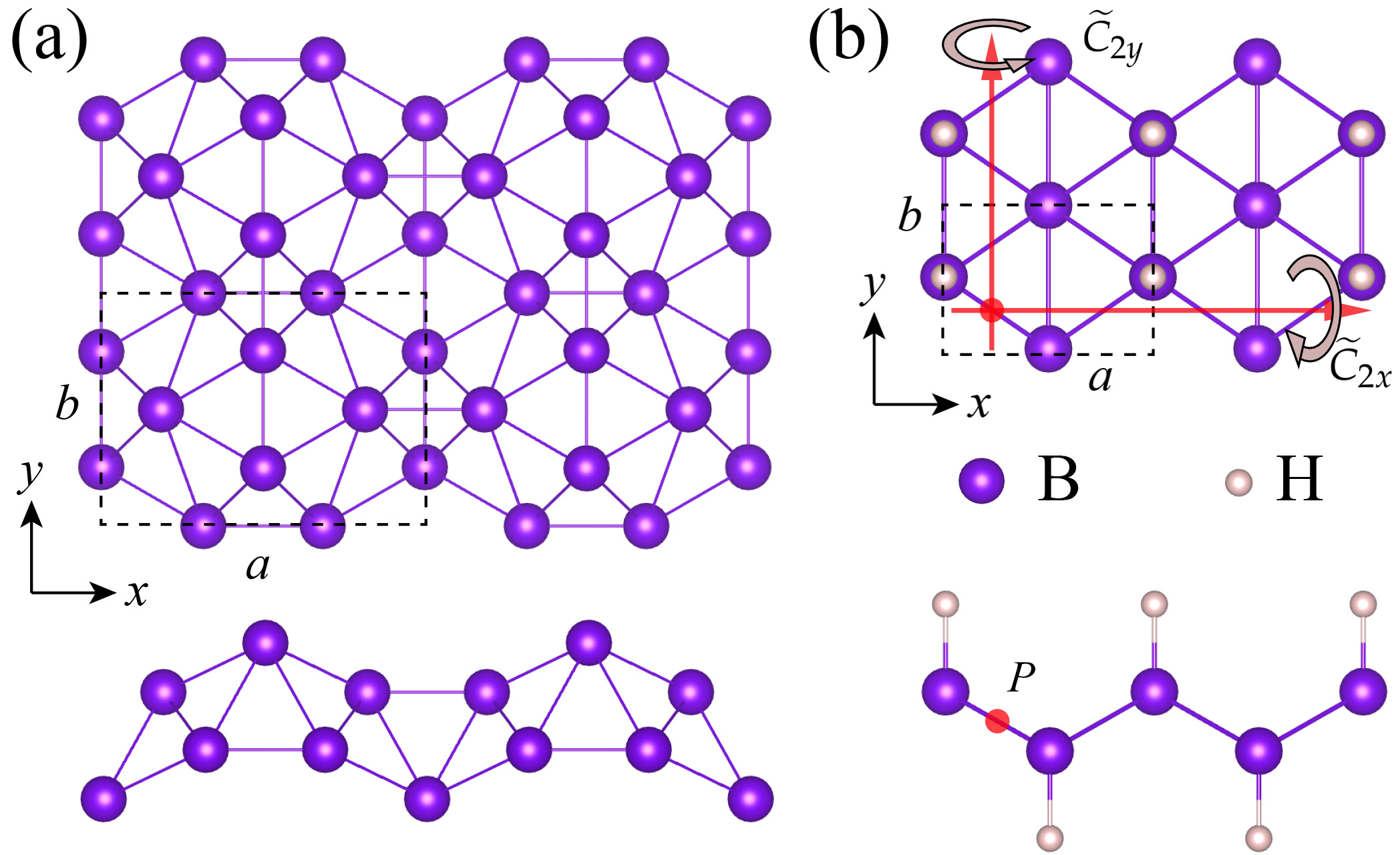}
\caption{(a)-(b) The top view and side view of 2D single layer B$_8$ and B$_2$H$_2$. The primitive unit cells are highlighted by the black dashed rectangular in each panel. The red arrows indicate the screw axes $\widetilde{C}_{2x}$ and $\widetilde{C}_{2y}$, respectively. The red dot represents the inversion center.
}\label{structure}
\end{figure}

The geometric structures of B$_8$ and B$_{2}$H$_{2}$ are displayed in Fig.~\ref{structure}. As shown in Fig.~\ref{structure}(a), there are two inequivalent boron atoms in the unit cell, one is located at Wyckoff position 4f(0.315, 0.000, 0.480) and the other at 4e(0.000, 0.253, 0.445). All boron atoms form an arch-like sheet.
For B$_{2}$H$_{2}$ in Fig.~\ref{structure}(b), there is a boron atom at Wyckoff position 2b(0.500,0.000,0.522) and a hydrogen atom at 2b(0.500,0.000,0.588). It can be regarded as a 2D borophene sheet, which has been experimentally synthesized\cite{borophenesy}, passivated with hydrogen atoms from each side alternately.

Both crystal structures have two screw axes: ${{\widetilde{C}}_{2x}}=\{{{C}_{2x}}|a/2\}$ and ${{\widetilde{C}}_{2y}}=\{{{C}_{2y}}|b/2\}$ as displayed in Fig.~\ref{structure}(b). The inversion center locates at the crossing point of two screw axes.
The fully optimized lattice parameters are $a=4.523~\mathring{\mathrm{A}}$, $b=3.260~\mathring{\mathrm{A}}$ for B$_8$ and $a=2.822~\mathring{\mathrm{A}}$, $b=1.937~\mathring{\mathrm{A}}$ for B$_2$H$_2$, respectively.

The first-principles calculations are performed based on the density functional theory (DFT) by using the Vienna ab initio simulation package (VASP)\cite{28} with projector augmented wave (PAW)\cite{32}. The exchange-correlation part is described with the generalized gradient approximation (GGA)\cite{33} in the scheme of Perdew-Burke-Ernzerhof (PBE) functional\cite{29}. To confirm the convergence of calculations, the kinetic cutoff energy is set as $450~\mathrm{eV}$, and a $\Gamma$ centered k-mesh of $15 \times 11 \times 1$ is employed.
The lattice vector along $c$ direction is chosen as $20.0~\mathring{\mathrm{A}}$, which is large enough to avoid interactions between adjacent layers. The maximally localized Wannier functions (MLWF) are obtained by Wannier90 code\cite{30,31}.

\subsection{Cat's cradle-like bandstuctures}

\begin{figure}
\includegraphics[width=3.5 in]{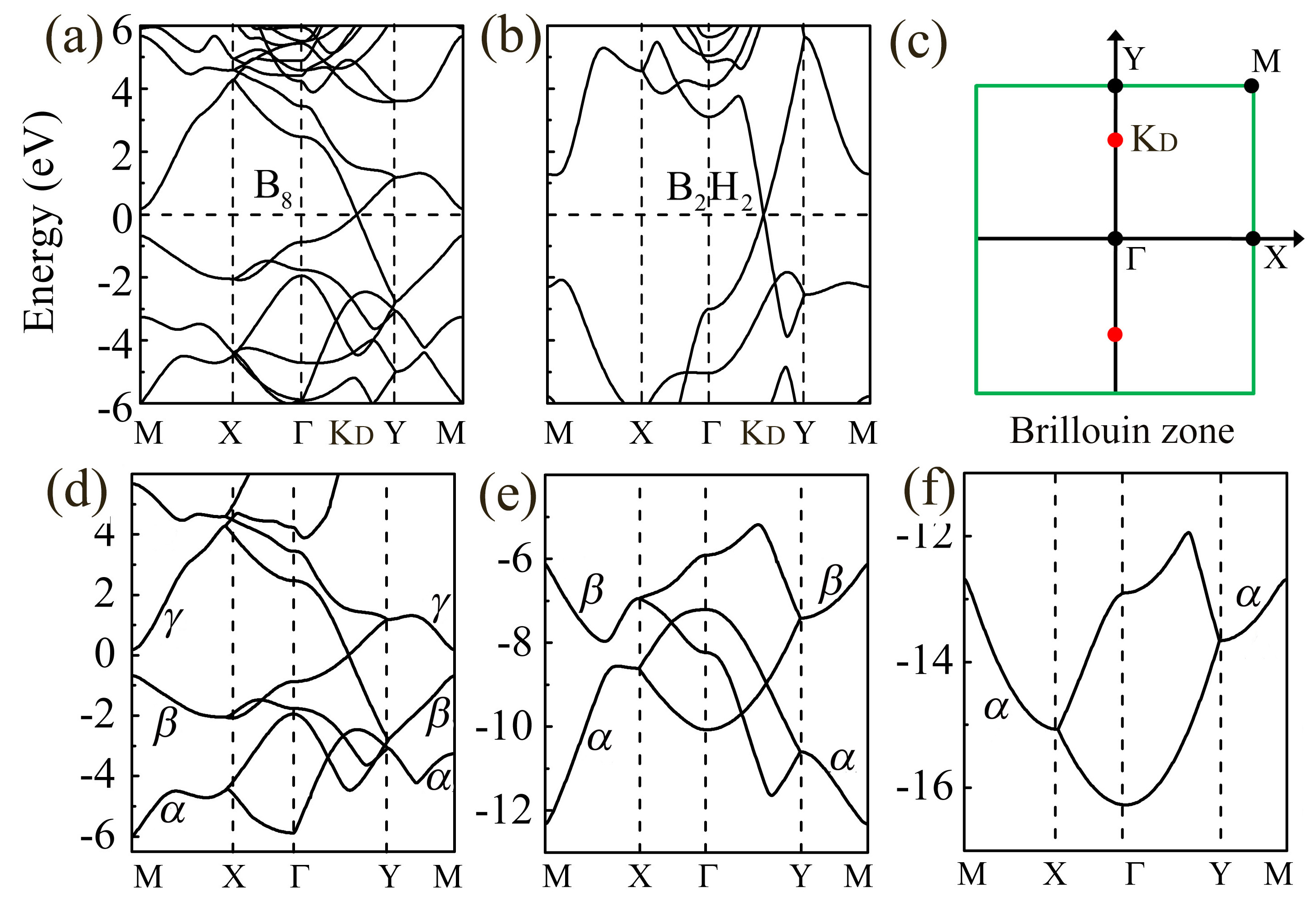}
\caption{(color online).(a)-(b) The electronic band structures of B$_{8}$ and B$_{2}$H$_{2}$, respectively. The Fermi level is set to be zero. (c) The first Brillouin zone. The green line represents double-degenerate k-points. The red points stand for location of the Dirac point at Fermi level, labeled by $K_D$. (d)-(f) The highlighted band structures for B$_{8}$.
}\label{dftband1}
\end{figure}

The band structures of monolayer B$_8$ and B$_2$H$_2$ are shown in Fig.~\ref{dftband1}(a) and (b).
We can see two common features of those bandstructures. Firstly, all bands are double-degenerate along the BZ boundary (e.g. XM and YM paths) and come into pairs. Secondly, all the double-degenerate bands are splitting inside BZ.
Those two features result from two screw rotation symmetries as discussed in part II.
What's more interesting is the emergence of Dirac points along $\Gamma$Y path at the Fermi level, which belongs to the type-I case in Table~\ref{Table.str}.
Taking B$_8$ case for example, as shown in Fig.~\ref{dftband1}(d), one band from pair $\beta$  inverted with one band from pair $\gamma$ around $\Gamma$ point that gives birth to the Dirac points at Fermi level. Meanwhile, band inversion also happens between pair $\beta$ and pair $\alpha$  which in fact couples three pair of bands together and forms a cat's cradle-like Dirac semimetal state as depicted in Fig.~\ref{band1}(c).
In Fig.~\ref{dftband1}(e), within the energy range between -13 eV and -5 eV, we discover two pairs of bands ($\alpha$ and $\beta$) are inverted forming an hourglass-like structure as predicted in Fig.~\ref{band1}(b).
In Fig.~\ref{dftband1}(f), we find a pair of lowest valence bands isolated from all other bands, which belongs to the case shown in Fig.~\ref{band1}(a).

For two Dirac points around the Fermi level, utilizing the $k{\cdot}p$ model given in Eq.~(\ref{Hkp1}), we expand the Hamiltonian to the linear term around two Dirac points $K_{D}=(0, \pm \sqrt{m_{0}/m_{2}}$) and obtain the low-energy effective $k{\cdot}p$ model as following:
\begin{eqnarray}
H_{eff}^{\kappa}=\kappa w{_y}k_{y}\tau_{0}+u_{x}k_{x}\tau_{y}+\kappa u_{y}k_{y}\tau_{z}. \label{Heff}
\end{eqnarray}
The $\kappa=\pm1$ is the valley index, $u_{x,y}$ are Fermi velocities near the Dirac points. The $w_{y}k_{y}\sigma_0$ describes the tilt term, which can give type-II Dirac cone if $w_{y}$ is larger than $u_{y}$ as what happens in monolayer WTe$_2$.
By fitting $k{\cdot}p$ model with DFT result, we get these velocities as $u_x$=5.2$\times$$10^5$ m/s, $u_y$=8.0$\times$$10^5$ m/s, $w_y$=3.4$\times$$10^5$ m/s for B$_{8}$, and $u_x$=7.7$\times$$10^5$ m/s, $u_y$=13.9$\times$$10^5$ m/s, $w_y$=3.5$\times$$10^5$ m/s for B$_{2}$H$_{2}$. Obviously, a remarkable anisotropy effect can be realized in both of this system, which provides a platform to study the intriguing physical properties related to the anisotropy.

\subsection{Topological properties and edge states}

\begin{figure}[t]
\includegraphics[width=8.5 cm]{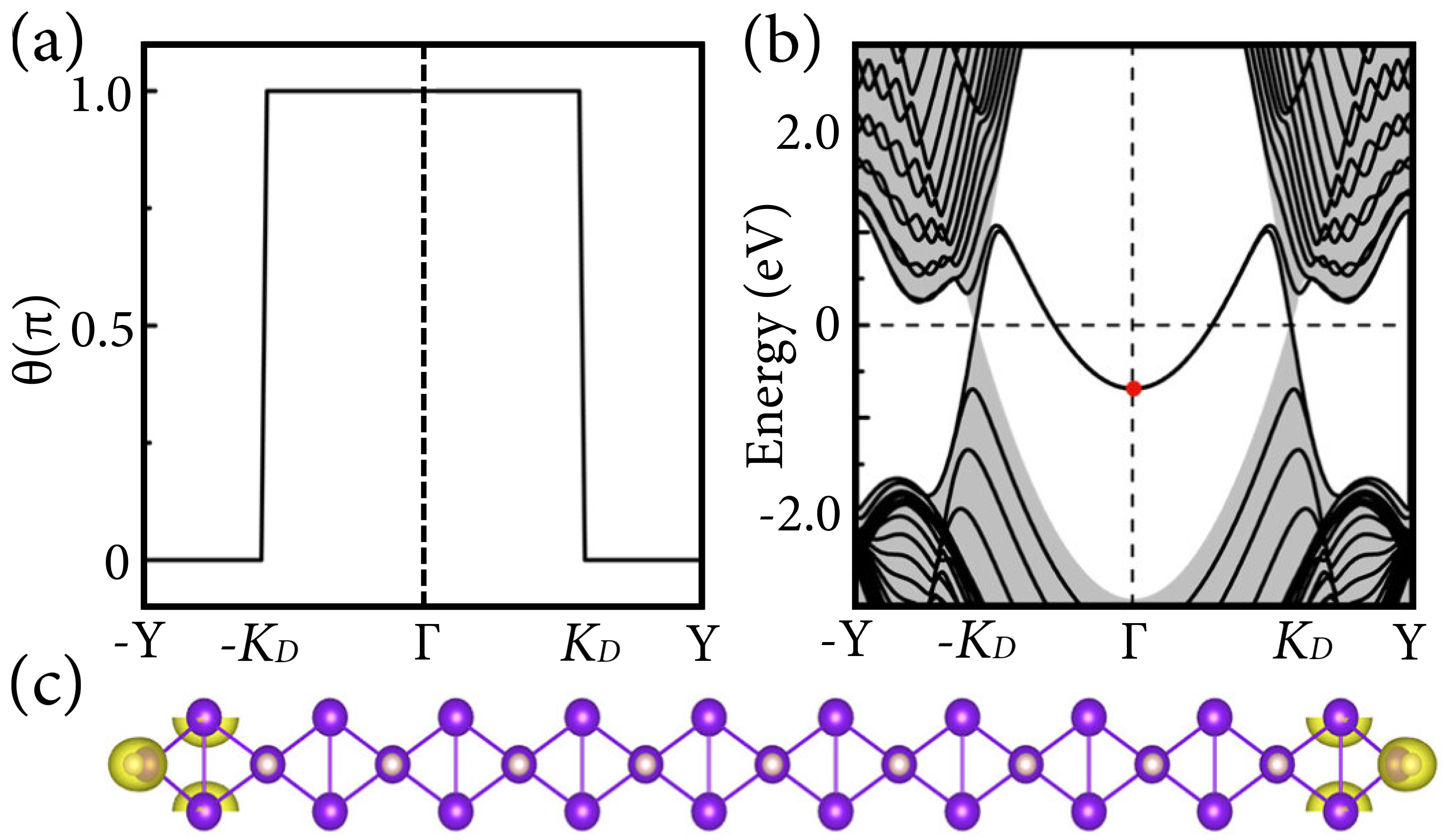}
\caption{(color online).(a) The $k$-dependent Berry phase for B$_{2}$H$_{2}$. (b) The band structure of B$_{2}$H$_{2}$ nanoribbon with width of 28.50 {\AA}. The gray color represents the bulk states for an infinite system. (c) The charge density distribution of edge state at $\Gamma$ point in (b).
}\label{ribbon}
\end{figure}

Above we have shown the nonsymmorphic symmetry protected cat's cradle-like Dirac semimetal in B$_{8}$ and B$_{2}$H$_{2}$. Here we further reveal its topology nature by defining a $k$-dependent Berry phase and demonstrating corresponding nontrivial edge states.
The Berry phase is defined as:
\begin{eqnarray}
\theta\left(k_{y}\right)=-i\sum_{n\in Occ}\int_{-\pi}^{\pi}d{k_{x}}\left\langle u_{n}\left(\boldsymbol{k}\right)\left|\partial_{k_{x}}\right|u_{n}\left(\boldsymbol{k}\right)\right\rangle  \label{berry1}
\end{eqnarray}
where  $u_{n}\left(\boldsymbol{k}\right)$ is the periodic component of Bloch wave functions and the integral takes for all the occupied bands.
With symmetry constraint, the Berry phase must be quantized to 0 or $\pi$ modulo 2$\pi$\cite{35} for any screw-symmetric momentum loops.
As shown in Fig.~\ref{ribbon}(a), we do the integral in Eq.~(\ref{berry1}) along $k_{x}$ direction and get $k_{y}$-dependent Berry phase. We can see the Berry phase is $\pi$ for any $k_{y}$ in the region of [$-K_D$, $K_D$] and is zero outside this region. Comparing it with the band structure, we find the jump of Berry phase is accompanied with the gap closing and reopening process along $k_{y}$ direction. The nonzero Berry phase indicates charge pumping \cite{36} at the edge of the system. From those perspectives, the Dirac point in 2D BZ is analogous with the Weyl points in 3D BZ where the nonzero Berry phase (Chern number) appears for the integral over any 2D plane between a pair of Weyl points with opposite charities.

Like in the 3D Weyl semimetals, the existence of topological edge states or Fermi arc is expected in the 2D Dirac semimetals.
In Fig.~\ref{ribbon}(c), we build a nanoribbon of B$_{2}$H$_{2}$ with width of 28.50 {\AA}, which is periodic only in $y$ direction.
The band structure is displayed in Fig.~\ref{ribbon}(b), where we can see a pair of Dirac points in the bulk are projected in one dimension BZ and the topological edge state emerges from one Dirac point to the other Dirac point inside region [$-K_D$, $K_D$] with nontrivial $\pi$ Berry phase.
In order to demonstrate the distribution of the edge state, we plot its charge density in real space in Fig.~\ref{ribbon}(c), which is well-localized at the edge atoms.
This edge state originates from the topological nature of Dirac point and is robust against external perturbation that can provide an ideal conducting channel in transport.

\subsection{Response to the strain and electric field}
\begin{figure}
\includegraphics[width=8.5 cm]{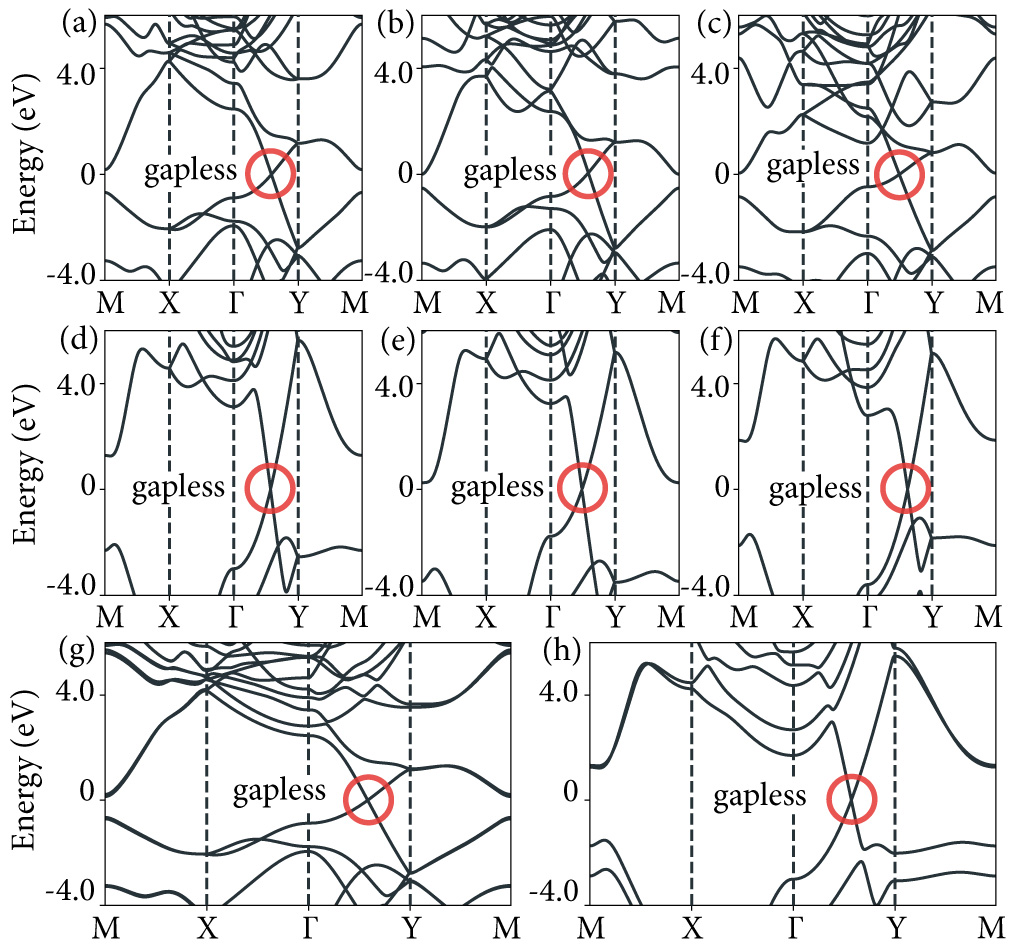}
\caption{(color online). (a)-(c) The electronic band structures of B$_{8}$ under biaxial strain, uniaxial strain along $b$ and $a$ direction. The strain strength is taken as 10\%. (d)-(f) The electronic band structures of B$_{2}$H$_{2}$ under the same biaxial and uniaxial strain along $b$ and $a$ directions.
(g)-(h) The electronic band structure of B$_{8}$ and B$_{2}$H$_{2}$ under vertical E-field of 0.5 eV/{\AA}, respectively.
}\label{strain}
\end{figure}

To test the symmetry protected nature of Dirac cone in cat's cradle-like Dirac semimetal, we impose both biaxial strain and uniaxial strains along $a$ and $b$ directions for B$_8$ and B$_{2}$H$_{2}$.
Under 10$\%$ strains, we can see the Dirac cone at Fermi level and band degeneracy along the BZ boundary are nearly unchanged. Thus, the strains can not change the cat's cradle-like bandstructure. However, we find that the position of Dirac point slightly moves toward Y ($\Gamma$) when the extensive strain is applying extensive uniaxial strain along $a$ ($b$) axis.
On the other hand, by applying a vertical electric field, the inversion symmetry (${P}$) and screw symmetries ($\widetilde{C}_{2x}$,$\widetilde{C}_{2y}$) are broken while the mirror symmetries ($\widetilde{M}_{x}$, $\widetilde{M}_{y}$) are remained.
Thus, the degenerate bands at the BZ boundary are splitting as shown in Fig.~\ref{strain}(g),(h).
Nevertheless, as we have demonstrated that the Dirac point is individually protected by any one of the $\widetilde{C}_{2y}$, $\widetilde{M}_{x}$, so the Dirac cone at Fermi level is still reserved because of $\widetilde{M}_{x}$. From this perspective, the Dirac cones in B$_{8}$ and B$_{2}$H$_{2}$ are very robust against external strain and electric field that may provide excellent transport performance in experiment.

In the end, we consider the impact of SOC effect, and find there is almost no visible difference between the band structures with and without SOC except at Dirac point where a extremely tiny gap is induced by SOC effects that make the systems to be topological insulators.
The nontrivial gaps opened by SOC are 0.03 meV  and 2.25 meV for B$_{8}$ and B$_{2}$H$_{2}$, respectively, which is negligible in experiment.

\section{Conclusion}
Recent experimental discovery of Dirac semimetals in 2D materials beyond graphene, such as few-layer black phosphorus\cite{kim2015observation} and ${\beta_{12}}$/${\chi_{3}}$-borophene\cite{fengprl,feng2018discovery}, have attracted extensive attentions. In those Dirac semimetal materials, the symmorphic symmetry like space inversion or mirror reflection symmetry plays the crucial role. On the other hand, the monolayer WTe$_{2}$ is presented as the first example of nonsymmorphic semimetals protected by one screw axis in the absence of SOC, which is diagnosed by a topological invariant associated with a non-Abelian Berry gauge field. However, the nonnegligible SOC of WTe$_{2}$ make it a significant topological insulator\cite{qian2014quantum,wu2018observation}, which can realize quantum spin Hall effect at the temperature of 100K. In this paper, we have expanded the nonsymmorphic semimetal from one screw axis to multiple screw axes, and we found the novel cat's cradle-like Dirac semimetal which is protected by certain nonsymmorphic layer groups with multiple screw axes and originated from band inversion mechanism. Such novel topological semimetal phase is first proposed in monolayer B$_{2}$H$_{2}$(B$_{8}$) which is possibly synthesized by hydrogenation of 2D borophene sheet. Finally, we demonstrate its topological properties by the calculation of quantized Berry phase along the symmetric-loop and flat Fermi-arc edge states of nanoribbons. We expect that the new nonsymmorphic semimetal with cat's cradle-like band structure can give fascinating properties in future experiment.

\begin{acknowledgments}
This work was supported by the National Natural Science Foundation of China (Nos. 11774028, 11734003, 11574029, 11404022), the National Key R\&D Program of China (No. 2016YFA0300600),
the MOST Project of China (No. 2014CB920903),  and Basic Research Funds of Beijing Institute of Technology (No. 2017CX01018).
\end{acknowledgments}

\bibliography{ref}

\begin{thebibliography}{52}%
\makeatletter
\providecommand \@ifxundefined [1]{%
 \@ifx{#1\undefined}
}%
\providecommand \@ifnum [1]{%
 \ifnum #1\expandafter \@firstoftwo
 \else \expandafter \@secondoftwo
 \fi
}%
\providecommand \@ifx [1]{%
 \ifx #1\expandafter \@firstoftwo
 \else \expandafter \@secondoftwo
 \fi
}%
\providecommand \natexlab [1]{#1}%
\providecommand \enquote  [1]{``#1''}%
\providecommand \bibnamefont  [1]{#1}%
\providecommand \bibfnamefont [1]{#1}%
\providecommand \citenamefont [1]{#1}%
\providecommand \href@noop [0]{\@secondoftwo}%
\providecommand \href [0]{\begingroup \@sanitize@url \@href}%
\providecommand \@href[1]{\@@startlink{#1}\@@href}%
\providecommand \@@href[1]{\endgroup#1\@@endlink}%
\providecommand \@sanitize@url [0]{\catcode `\\12\catcode `\$12\catcode
  `\&12\catcode `\#12\catcode `\^12\catcode `\_12\catcode `\%12\relax}%
\providecommand \@@startlink[1]{}%
\providecommand \@@endlink[0]{}%
\providecommand \url  [0]{\begingroup\@sanitize@url \@url }%
\providecommand \@url [1]{\endgroup\@href {#1}{\urlprefix }}%
\providecommand \urlprefix  [0]{URL }%
\providecommand \Eprint [0]{\href }%
\providecommand \doibase [0]{http://dx.doi.org/}%
\providecommand \selectlanguage [0]{\@gobble}%
\providecommand \bibinfo  [0]{\@secondoftwo}%
\providecommand \bibfield  [0]{\@secondoftwo}%
\providecommand \translation [1]{[#1]}%
\providecommand \BibitemOpen [0]{}%
\providecommand \bibitemStop [0]{}%
\providecommand \bibitemNoStop [0]{.\EOS\space}%
\providecommand \EOS [0]{\spacefactor3000\relax}%
\providecommand \BibitemShut  [1]{\csname bibitem#1\endcsname}%
\let\auto@bib@innerbib\@empty
\bibitem [{\citenamefont {Castro~Neto}\ \emph {et~al.}(2009)\citenamefont
  {Castro~Neto}, \citenamefont {Guinea}, \citenamefont {Peres}, \citenamefont
  {Novoselov},\ and\ \citenamefont {Geim}}]{RMPgra}%
  \BibitemOpen
  \bibfield  {author} {\bibinfo {author} {\bibfnamefont {A.~H.}\ \bibnamefont
  {Castro~Neto}}, \bibinfo {author} {\bibfnamefont {F.}~\bibnamefont {Guinea}},
  \bibinfo {author} {\bibfnamefont {N.~M.~R.}\ \bibnamefont {Peres}}, \bibinfo
  {author} {\bibfnamefont {K.~S.}\ \bibnamefont {Novoselov}}, \ and\ \bibinfo
  {author} {\bibfnamefont {A.~K.}\ \bibnamefont {Geim}},\ }\href {\doibase
  10.1103/RevModPhys.81.109} {\bibfield  {journal} {\bibinfo  {journal} {Rev.
  Mod. Phys.}\ }\textbf {\bibinfo {volume} {81}},\ \bibinfo {pages} {109}
  (\bibinfo {year} {2009})}\BibitemShut {NoStop}%
\bibitem [{\citenamefont {Ponomarenko}\ \emph {et~al.}(2009)\citenamefont
  {Ponomarenko}, \citenamefont {Yang}, \citenamefont {Mohiuddin}, \citenamefont
  {Katsnelson}, \citenamefont {Novoselov}, \citenamefont {Morozov},
  \citenamefont {Zhukov}, \citenamefont {Schedin}, \citenamefont {Hill},\ and\
  \citenamefont {Geim}}]{1}%
  \BibitemOpen
  \bibfield  {author} {\bibinfo {author} {\bibfnamefont {L.~A.}\ \bibnamefont
  {Ponomarenko}}, \bibinfo {author} {\bibfnamefont {R.}~\bibnamefont {Yang}},
  \bibinfo {author} {\bibfnamefont {T.~M.}\ \bibnamefont {Mohiuddin}}, \bibinfo
  {author} {\bibfnamefont {M.~I.}\ \bibnamefont {Katsnelson}}, \bibinfo
  {author} {\bibfnamefont {K.~S.}\ \bibnamefont {Novoselov}}, \bibinfo {author}
  {\bibfnamefont {S.~V.}\ \bibnamefont {Morozov}}, \bibinfo {author}
  {\bibfnamefont {A.~A.}\ \bibnamefont {Zhukov}}, \bibinfo {author}
  {\bibfnamefont {F.}~\bibnamefont {Schedin}}, \bibinfo {author} {\bibfnamefont
  {E.~W.}\ \bibnamefont {Hill}}, \ and\ \bibinfo {author} {\bibfnamefont
  {A.~K.}\ \bibnamefont {Geim}},\ }\href {\doibase
  10.1103/PhysRevLett.102.206603} {\bibfield  {journal} {\bibinfo  {journal}
  {Phys. Rev. Lett.}\ }\textbf {\bibinfo {volume} {102}},\ \bibinfo {pages}
  {206603} (\bibinfo {year} {2009})}\BibitemShut {NoStop}%
\bibitem [{\citenamefont {Novoselov}\ \emph {et~al.}(2005)\citenamefont
  {Novoselov}, \citenamefont {Geim}, \citenamefont {Morozov}, \citenamefont
  {Jiang}, \citenamefont {Katsnelson}, \citenamefont {Grigorieva},
  \citenamefont {Dubonos},\ and\ \citenamefont {Firsov}}]{2}%
  \BibitemOpen
  \bibfield  {author} {\bibinfo {author} {\bibfnamefont {K.~S.}\ \bibnamefont
  {Novoselov}}, \bibinfo {author} {\bibfnamefont {A.~K.}\ \bibnamefont {Geim}},
  \bibinfo {author} {\bibfnamefont {S.~V.}\ \bibnamefont {Morozov}}, \bibinfo
  {author} {\bibfnamefont {D.}~\bibnamefont {Jiang}}, \bibinfo {author}
  {\bibfnamefont {M.~I.}\ \bibnamefont {Katsnelson}}, \bibinfo {author}
  {\bibfnamefont {I.~V.}\ \bibnamefont {Grigorieva}}, \bibinfo {author}
  {\bibfnamefont {S.~V.}\ \bibnamefont {Dubonos}}, \ and\ \bibinfo {author}
  {\bibfnamefont {A.~A.}\ \bibnamefont {Firsov}},\ }\href
  {http://dx.doi.org/10.1038/nature04233} {\bibfield  {journal} {\bibinfo
  {journal} {Nature}\ }\textbf {\bibinfo {volume} {438}},\ \bibinfo {pages}
  {197} (\bibinfo {year} {2005})}\BibitemShut {NoStop}%
\bibitem [{\citenamefont {Zhang}\ \emph {et~al.}(2005)\citenamefont {Zhang},
  \citenamefont {Tan}, \citenamefont {Stormer},\ and\ \citenamefont {Kim}}]{3}%
  \BibitemOpen
  \bibfield  {author} {\bibinfo {author} {\bibfnamefont {Y.}~\bibnamefont
  {Zhang}}, \bibinfo {author} {\bibfnamefont {Y.-W.}\ \bibnamefont {Tan}},
  \bibinfo {author} {\bibfnamefont {H.~L.}\ \bibnamefont {Stormer}}, \ and\
  \bibinfo {author} {\bibfnamefont {P.}~\bibnamefont {Kim}},\ }\href {\doibase
  10.1080/00018732.2014.927109} {\bibfield  {journal} {\bibinfo  {journal}
  {Nature}\ }\textbf {\bibinfo {volume} {438}},\ \bibinfo {pages} {201}
  (\bibinfo {year} {2005})}\BibitemShut {NoStop}%
\bibitem [{\citenamefont {Weiss}\ \emph {et~al.}(2012)\citenamefont {Weiss},
  \citenamefont {Zhou}, \citenamefont {Liao}, \citenamefont {Liu},
  \citenamefont {Jiang}, \citenamefont {Huang},\ and\ \citenamefont
  {Duan}}]{5}%
  \BibitemOpen
  \bibfield  {author} {\bibinfo {author} {\bibfnamefont {N.~O.}\ \bibnamefont
  {Weiss}}, \bibinfo {author} {\bibfnamefont {H.}~\bibnamefont {Zhou}},
  \bibinfo {author} {\bibfnamefont {L.}~\bibnamefont {Liao}}, \bibinfo {author}
  {\bibfnamefont {Y.}~\bibnamefont {Liu}}, \bibinfo {author} {\bibfnamefont
  {S.}~\bibnamefont {Jiang}}, \bibinfo {author} {\bibfnamefont
  {Y.}~\bibnamefont {Huang}}, \ and\ \bibinfo {author} {\bibfnamefont
  {X.}~\bibnamefont {Duan}},\ }\href {https://doi.org/10.1002/adma.201290269}
  {\bibfield  {journal} {\bibinfo  {journal} {Adv. Mater.}\ }\textbf {\bibinfo
  {volume} {24}},\ \bibinfo {pages} {5776} (\bibinfo {year}
  {2012})}\BibitemShut {NoStop}%
\bibitem [{\citenamefont {Burkov}\ \emph {et~al.}(2011)\citenamefont {Burkov},
  \citenamefont {Hook},\ and\ \citenamefont {Balents}}]{DNL1}%
  \BibitemOpen
  \bibfield  {author} {\bibinfo {author} {\bibfnamefont {A.A.}~\bibnamefont
  {Burkov}}, \bibinfo {author} {\bibfnamefont {M.D.}~\bibnamefont {Hook}}, \ and\
  \bibinfo {author} {\bibfnamefont {L.}~\bibnamefont {Balents}},\ }\href
  {https://journals.aps.org/prb/abstract/10.1103/PhysRevB.84.235126} {\bibfield
   {journal} {\bibinfo  {journal} {Phys. Rev. B}\ }\textbf {\bibinfo {volume}
  {84}},\ \bibinfo {pages} {235126} (\bibinfo {year} {2011})}\BibitemShut
  {NoStop}%
\bibitem [{\citenamefont {Fang}\ \emph {et~al.}(2016)\citenamefont {Fang},
  \citenamefont {Weng}, \citenamefont {Dai},\ and\ \citenamefont
  {Fang}}]{DNL2}%
  \BibitemOpen
  \bibfield  {author} {\bibinfo {author} {\bibfnamefont {C.}~\bibnamefont
  {Fang}}, \bibinfo {author} {\bibfnamefont {H.}~\bibnamefont {Weng}}, \bibinfo
  {author} {\bibfnamefont {X.}~\bibnamefont {Dai}}, \ and\ \bibinfo {author}
  {\bibfnamefont {Z.}~\bibnamefont {Fang}},\ }\href
  {http://iopscience.iop.org/article/10.1088/1674-1056/25/11/117106/meta}
  {\bibfield  {journal} {\bibinfo  {journal} {Chin. Phys. B}\ }\textbf
  {\bibinfo {volume} {25}},\ \bibinfo {pages} {117106} (\bibinfo {year}
  {2016})}\BibitemShut {NoStop}%
\bibitem [{\citenamefont {Young}\ and\ \citenamefont {Kane}(2015)}]{22}%
  \BibitemOpen
  \bibfield  {author} {\bibinfo {author} {\bibfnamefont {S.~M.}\ \bibnamefont
  {Young}}\ and\ \bibinfo {author} {\bibfnamefont {C.~L.}\ \bibnamefont
  {Kane}},\ }\href@noop {} {\bibfield  {journal} {\bibinfo  {journal} {Phys.
  Rev. Lett.}\ }\textbf {\bibinfo {volume} {115}},\ \bibinfo {pages} {126803}
  (\bibinfo {year} {2015})}\BibitemShut {NoStop}%
\bibitem [{\citenamefont {Fang}\ \emph {et~al.}(2012)\citenamefont {Fang},
  \citenamefont {Gilbert}, \citenamefont {Dai},\ and\ \citenamefont
  {Bernevig}}]{fang2012multi}%
  \BibitemOpen
  \bibfield  {author} {\bibinfo {author} {\bibfnamefont {C.}~\bibnamefont
  {Fang}}, \bibinfo {author} {\bibfnamefont {M.~J.}\ \bibnamefont {Gilbert}},
  \bibinfo {author} {\bibfnamefont {X.}~\bibnamefont {Dai}}, \ and\ \bibinfo
  {author} {\bibfnamefont {B.~A.}\ \bibnamefont {Bernevig}},\ }\href@noop {}
  {\bibfield  {journal} {\bibinfo  {journal} {Physical review letters}\
  }\textbf {\bibinfo {volume} {108}},\ \bibinfo {pages} {266802} (\bibinfo
  {year} {2012})}\BibitemShut {NoStop}%
\bibitem [{\citenamefont {Wang}\ \emph {et~al.}(2016)\citenamefont {Wang},
  \citenamefont {Alexandradinata}, \citenamefont {Cava},\ and\ \citenamefont
  {Bernevig}}]{wang2016hourglass}%
  \BibitemOpen
  \bibfield  {author} {\bibinfo {author} {\bibfnamefont {Z.}~\bibnamefont
  {Wang}}, \bibinfo {author} {\bibfnamefont {A.}~\bibnamefont
  {Alexandradinata}}, \bibinfo {author} {\bibfnamefont {R.~J.}\ \bibnamefont
  {Cava}}, \ and\ \bibinfo {author} {\bibfnamefont {B.~A.}\ \bibnamefont
  {Bernevig}},\ }\href {https://www.nature.com/articles/nature17410} {\bibfield
   {journal} {\bibinfo  {journal} {Nature}\ }\textbf {\bibinfo {volume}
  {532}},\ \bibinfo {pages} {189} (\bibinfo {year} {2016})}\BibitemShut
  {NoStop}%
\bibitem [{\citenamefont {Po}\ \emph {et~al.}(2017)\citenamefont {Po},
  \citenamefont {Vishwanath},\ and\ \citenamefont {Watanabe}}]{po2017symmetry}%
  \BibitemOpen
  \bibfield  {author} {\bibinfo {author} {\bibfnamefont {H.~C.}\ \bibnamefont
  {Po}}, \bibinfo {author} {\bibfnamefont {A.}~\bibnamefont {Vishwanath}}, \
  and\ \bibinfo {author} {\bibfnamefont {H.}~\bibnamefont {Watanabe}},\ }\href
  {https://www.nature.com/articles/s41467-017-00133-2} {\bibfield  {journal}
  {\bibinfo  {journal} {Nature Communications}\ }\textbf {\bibinfo {volume}
  {8}},\ \bibinfo {pages} {50} (\bibinfo {year} {2017})}\BibitemShut {NoStop}%
\bibitem [{\citenamefont {van Miert}\ and\ \citenamefont {Smith}(2016)}]{8}%
  \BibitemOpen
  \bibfield  {author} {\bibinfo {author} {\bibfnamefont {G.}~\bibnamefont {van
  Miert}}\ and\ \bibinfo {author} {\bibfnamefont {C.~M.}\ \bibnamefont
  {Smith}},\ }\href {\doibase 10.1103/PhysRevB.93.035401} {\bibfield  {journal}
  {\bibinfo  {journal} {Phys. Rev. B}\ }\textbf {\bibinfo {volume} {93}},\
  \bibinfo {pages} {035401} (\bibinfo {year} {2016})}\BibitemShut {NoStop}%
\bibitem [{\citenamefont {Weng}\ \emph {et~al.}(2015)\citenamefont {Weng},
  \citenamefont {Liang}, \citenamefont {Xu}, \citenamefont {Yu}, \citenamefont
  {Fang}, \citenamefont {Dai},\ and\ \citenamefont {Kawazoe}}]{PDNL1}%
  \BibitemOpen
  \bibfield  {author} {\bibinfo {author} {\bibfnamefont {H.}~\bibnamefont
  {Weng}}, \bibinfo {author} {\bibfnamefont {Y.}~\bibnamefont {Liang}},
  \bibinfo {author} {\bibfnamefont {Q.}~\bibnamefont {Xu}}, \bibinfo {author}
  {\bibfnamefont {R.}~\bibnamefont {Yu}}, \bibinfo {author} {\bibfnamefont
  {Z.}~\bibnamefont {Fang}}, \bibinfo {author} {\bibfnamefont {X.}~\bibnamefont
  {Dai}}, \ and\ \bibinfo {author} {\bibfnamefont {Y.}~\bibnamefont
  {Kawazoe}},\ }\href@noop {} {\bibfield  {journal} {\bibinfo  {journal} {Phys.
  Rev. B}\ }\textbf {\bibinfo {volume} {92}},\ \bibinfo {pages} {045108}
  (\bibinfo {year} {2015})}\BibitemShut {NoStop}%
\bibitem [{\citenamefont {Yu}\ \emph {et~al.}(2015)\citenamefont {Yu},
  \citenamefont {Weng}, \citenamefont {Fang}, \citenamefont {Dai},\ and\
  \citenamefont {Hu}}]{PDNL2}%
  \BibitemOpen
  \bibfield  {author} {\bibinfo {author} {\bibfnamefont {R.}~\bibnamefont
  {Yu}}, \bibinfo {author} {\bibfnamefont {H.}~\bibnamefont {Weng}}, \bibinfo
  {author} {\bibfnamefont {Z.}~\bibnamefont {Fang}}, \bibinfo {author}
  {\bibfnamefont {X.}~\bibnamefont {Dai}}, \ and\ \bibinfo {author}
  {\bibfnamefont {X.}~\bibnamefont {Hu}},\ }\href {\doibase
  10.1103/PhysRevLett.115.036807} {\bibfield  {journal} {\bibinfo  {journal}
  {Phys. Rev. Lett.}\ }\textbf {\bibinfo {volume} {115}},\ \bibinfo {pages}
  {036807} (\bibinfo {year} {2015})}\BibitemShut {NoStop}%
\bibitem [{\citenamefont {Kim}\ \emph {et~al.}(2015{\natexlab{a}})\citenamefont
  {Kim}, \citenamefont {Wieder}, \citenamefont {Kane},\ and\ \citenamefont
  {Rappe}}]{17}%
  \BibitemOpen
  \bibfield  {author} {\bibinfo {author} {\bibfnamefont {Y.}~\bibnamefont
  {Kim}}, \bibinfo {author} {\bibfnamefont {B.~J.}\ \bibnamefont {Wieder}},
  \bibinfo {author} {\bibfnamefont {C.~L.}\ \bibnamefont {Kane}}, \ and\
  \bibinfo {author} {\bibfnamefont {A.~M.}\ \bibnamefont {Rappe}},\ }\href
  {\doibase 10.1103/PhysRevLett.115.036806} {\bibfield  {journal} {\bibinfo
  {journal} {Phys. Rev. Lett.}\ }\textbf {\bibinfo {volume} {115}},\ \bibinfo
  {pages} {036806} (\bibinfo {year} {2015}{\natexlab{a}})}\BibitemShut
  {NoStop}%
\bibitem [{\citenamefont {Feng}\ \emph
  {et~al.}(2017{\natexlab{a}})\citenamefont {Feng}, \citenamefont {Fu},
  \citenamefont {Kasamatsu}, \citenamefont {Ito}, \citenamefont {Cheng},
  \citenamefont {Liu}, \citenamefont {Feng}, \citenamefont {Wu}, \citenamefont
  {Mahatha}, \citenamefont {Sheverdyaeva} \emph {et~al.}}]{cusi}%
  \BibitemOpen
  \bibfield  {author} {\bibinfo {author} {\bibfnamefont {B.}~\bibnamefont
  {Feng}}, \bibinfo {author} {\bibfnamefont {B.}~\bibnamefont {Fu}}, \bibinfo
  {author} {\bibfnamefont {S.}~\bibnamefont {Kasamatsu}}, \bibinfo {author}
  {\bibfnamefont {S.}~\bibnamefont {Ito}}, \bibinfo {author} {\bibfnamefont
  {P.}~\bibnamefont {Cheng}}, \bibinfo {author} {\bibfnamefont {C.-C.}\
  \bibnamefont {Liu}}, \bibinfo {author} {\bibfnamefont {Y.}~\bibnamefont
  {Feng}}, \bibinfo {author} {\bibfnamefont {S.}~\bibnamefont {Wu}}, \bibinfo
  {author} {\bibfnamefont {S.~K.}\ \bibnamefont {Mahatha}}, \bibinfo {author}
  {\bibfnamefont {P.}~\bibnamefont {Sheverdyaeva}},  \emph {et~al.},\ }\href
  {https://www.nature.com/articles/s41467-017-01108-z} {\bibfield  {journal}
  {\bibinfo  {journal} {Nat. Commun.}\ }\textbf {\bibinfo {volume} {8}},\
  \bibinfo {pages} {1007} (\bibinfo {year} {2017}{\natexlab{a}})}\BibitemShut
  {NoStop}%
\bibitem [{\citenamefont {Chan}\ \emph {et~al.}(2016)\citenamefont {Chan},
  \citenamefont {Chiu}, \citenamefont {Chou},\ and\ \citenamefont
  {Schnyder}}]{19}%
  \BibitemOpen
  \bibfield  {author} {\bibinfo {author} {\bibfnamefont {Y.-H.}\ \bibnamefont
  {Chan}}, \bibinfo {author} {\bibfnamefont {C.-K.}\ \bibnamefont {Chiu}},
  \bibinfo {author} {\bibfnamefont {M.~Y.}\ \bibnamefont {Chou}}, \ and\
  \bibinfo {author} {\bibfnamefont {A.~P.}\ \bibnamefont {Schnyder}},\ }\href
  {\doibase 10.1103/PhysRevB.93.205132} {\bibfield  {journal} {\bibinfo
  {journal} {Phys. Rev. B}\ }\textbf {\bibinfo {volume} {93}},\ \bibinfo
  {pages} {205132} (\bibinfo {year} {2016})}\BibitemShut {NoStop}%
\bibitem [{\citenamefont {Bian}\ \emph {et~al.}(2016)\citenamefont {Bian},
  \citenamefont {Chang}, \citenamefont {Zheng}, \citenamefont {Velury},
  \citenamefont {Xu}, \citenamefont {Neupert}, \citenamefont {Chiu},
  \citenamefont {Huang}, \citenamefont {Sanchez}, \citenamefont {Belopolski},
  \citenamefont {Alidoust}, \citenamefont {Chen}, \citenamefont {Chang},
  \citenamefont {Bansil}, \citenamefont {Jeng}, \citenamefont {Lin},\ and\
  \citenamefont {Hasan}}]{20}%
  \BibitemOpen
  \bibfield  {author} {\bibinfo {author} {\bibfnamefont {G.}~\bibnamefont
  {Bian}}, \bibinfo {author} {\bibfnamefont {T.-R.}\ \bibnamefont {Chang}},
  \bibinfo {author} {\bibfnamefont {H.}~\bibnamefont {Zheng}}, \bibinfo
  {author} {\bibfnamefont {S.}~\bibnamefont {Velury}}, \bibinfo {author}
  {\bibfnamefont {S.-Y.}\ \bibnamefont {Xu}}, \bibinfo {author} {\bibfnamefont
  {T.}~\bibnamefont {Neupert}}, \bibinfo {author} {\bibfnamefont {C.-K.}\
  \bibnamefont {Chiu}}, \bibinfo {author} {\bibfnamefont {S.-M.}\ \bibnamefont
  {Huang}}, \bibinfo {author} {\bibfnamefont {D.~S.}\ \bibnamefont {Sanchez}},
  \bibinfo {author} {\bibfnamefont {I.}~\bibnamefont {Belopolski}}, \bibinfo
  {author} {\bibfnamefont {N.}~\bibnamefont {Alidoust}}, \bibinfo {author}
  {\bibfnamefont {P.-J.}\ \bibnamefont {Chen}}, \bibinfo {author}
  {\bibfnamefont {G.}~\bibnamefont {Chang}}, \bibinfo {author} {\bibfnamefont
  {A.}~\bibnamefont {Bansil}}, \bibinfo {author} {\bibfnamefont {H.-T.}\
  \bibnamefont {Jeng}}, \bibinfo {author} {\bibfnamefont {H.}~\bibnamefont
  {Lin}}, \ and\ \bibinfo {author} {\bibfnamefont {M.~Z.}\ \bibnamefont
  {Hasan}},\ }\href {\doibase 10.1103/PhysRevB.93.121113} {\bibfield  {journal}
  {\bibinfo  {journal} {Phys. Rev. B}\ }\textbf {\bibinfo {volume} {93}},\
  \bibinfo {pages} {121113} (\bibinfo {year} {2016})}\BibitemShut {NoStop}%
\bibitem [{\citenamefont {Michel}\ and\ \citenamefont {Zak}(1999)}]{21}%
  \BibitemOpen
  \bibfield  {author} {\bibinfo {author} {\bibfnamefont {L.}~\bibnamefont
  {Michel}}\ and\ \bibinfo {author} {\bibfnamefont {J.}~\bibnamefont {Zak}},\
  }\href {\doibase 10.1103/PhysRevB.59.5998} {\bibfield  {journal} {\bibinfo
  {journal} {Phys. Rev. B}\ }\textbf {\bibinfo {volume} {59}},\ \bibinfo
  {pages} {5998} (\bibinfo {year} {1999})}\BibitemShut {NoStop}%
\bibitem [{\citenamefont {Wang}(2017)}]{wangAFM2017}%
  \BibitemOpen
  \bibfield  {author} {\bibinfo {author} {\bibfnamefont {J.}~\bibnamefont
  {Wang}},\ }\href
  {https://journals.aps.org/prb/abstract/10.1103/PhysRevB.95.115138} {\bibfield
   {journal} {\bibinfo  {journal} {Phys. Rev. B}\ }\textbf {\bibinfo {volume}
  {95}},\ \bibinfo {pages} {115138} (\bibinfo {year} {2017})}\BibitemShut
  {NoStop}%
\bibitem [{\citenamefont {Wieder}\ and\ \citenamefont
  {Kane}(2016)}]{wiedercat2016}%
  \BibitemOpen
  \bibfield  {author} {\bibinfo {author} {\bibfnamefont {B.~J.}\ \bibnamefont
  {Wieder}}\ and\ \bibinfo {author} {\bibfnamefont {C.L.}~\bibnamefont {Kane}},\
  }\href {\doibase 10.1103/PhysRevB.94.155108} {\bibfield  {journal} {\bibinfo
  {journal} {Phys. Rev. B}\ }\textbf {\bibinfo {volume} {94}},\ \bibinfo
  {pages} {155108} (\bibinfo {year} {2016})}\BibitemShut {NoStop}%
\bibitem [{\citenamefont {Guan}\ \emph {et~al.}(2017)\citenamefont {Guan},
  \citenamefont {Liu}, \citenamefont {Yu}, \citenamefont {Wang}, \citenamefont
  {Yao},\ and\ \citenamefont {Yang}}]{23}%
  \BibitemOpen
  \bibfield  {author} {\bibinfo {author} {\bibfnamefont {S.}~\bibnamefont
  {Guan}}, \bibinfo {author} {\bibfnamefont {Y.}~\bibnamefont {Liu}}, \bibinfo
  {author} {\bibfnamefont {Z.-M.}\ \bibnamefont {Yu}}, \bibinfo {author}
  {\bibfnamefont {S.-S.}\ \bibnamefont {Wang}}, \bibinfo {author}
  {\bibfnamefont {Y.}~\bibnamefont {Yao}}, \ and\ \bibinfo {author}
  {\bibfnamefont {S.~A.}\ \bibnamefont {Yang}},\ }\href {\doibase
  10.1103/PhysRevMaterials.1.054003} {\bibfield  {journal} {\bibinfo  {journal}
  {Phys. Rev. Mater.}\ }\textbf {\bibinfo {volume} {1}},\ \bibinfo {pages}
  {054003} (\bibinfo {year} {2017})}\BibitemShut {NoStop}%
\bibitem [{\citenamefont {Muechler}\ \emph {et~al.}(2016)\citenamefont
  {Muechler}, \citenamefont {Alexandradinata}, \citenamefont {Neupert},\ and\
  \citenamefont {Car}}]{nonsm2016}%
  \BibitemOpen
  \bibfield  {author} {\bibinfo {author} {\bibfnamefont {L.}~\bibnamefont
  {Muechler}}, \bibinfo {author} {\bibfnamefont {A.}~\bibnamefont
  {Alexandradinata}}, \bibinfo {author} {\bibfnamefont {T.}~\bibnamefont
  {Neupert}}, \ and\ \bibinfo {author} {\bibfnamefont {R.}~\bibnamefont
  {Car}},\ }\href
  {https://journals.aps.org/prx/abstract/10.1103/PhysRevX.6.041069} {\bibfield
  {journal} {\bibinfo  {journal} {Phys. Rev. X}\ }\textbf {\bibinfo {volume}
  {6}},\ \bibinfo {pages} {041069} (\bibinfo {year} {2016})}\BibitemShut
  {NoStop}%
\bibitem [{\citenamefont {Zhou}\ \emph {et~al.}(2014)\citenamefont {Zhou},
  \citenamefont {Dong}, \citenamefont {Oganov}, \citenamefont {Zhu},
  \citenamefont {Tian},\ and\ \citenamefont {Wang}}]{zhouxfB8}%
  \BibitemOpen
  \bibfield  {author} {\bibinfo {author} {\bibfnamefont {X.-F.}\ \bibnamefont
  {Zhou}}, \bibinfo {author} {\bibfnamefont {X.}~\bibnamefont {Dong}}, \bibinfo
  {author} {\bibfnamefont {A.~R.}\ \bibnamefont {Oganov}}, \bibinfo {author}
  {\bibfnamefont {Q.}~\bibnamefont {Zhu}}, \bibinfo {author} {\bibfnamefont
  {Y.}~\bibnamefont {Tian}}, \ and\ \bibinfo {author} {\bibfnamefont {H.-T.}\
  \bibnamefont {Wang}},\ }\href {\doibase 10.1103/PhysRevLett.112.085502}
  {\bibfield  {journal} {\bibinfo  {journal} {Phys. Rev. Lett.}\ }\textbf
  {\bibinfo {volume} {112}},\ \bibinfo {pages} {085502} (\bibinfo {year}
  {2014})}\BibitemShut {NoStop}%
\bibitem [{\citenamefont {Jiao}\ \emph {et~al.}(2016)\citenamefont {Jiao},
  \citenamefont {Ma}, \citenamefont {Bell}, \citenamefont {Bilic},\ and\
  \citenamefont {Du}}]{jiaoB2H2}%
  \BibitemOpen
  \bibfield  {author} {\bibinfo {author} {\bibfnamefont {Y.}~\bibnamefont
  {Jiao}}, \bibinfo {author} {\bibfnamefont {F.}~\bibnamefont {Ma}}, \bibinfo
  {author} {\bibfnamefont {J.}~\bibnamefont {Bell}}, \bibinfo {author}
  {\bibfnamefont {A.}~\bibnamefont {Bilic}}, \ and\ \bibinfo {author}
  {\bibfnamefont {A.}~\bibnamefont {Du}},\ }\href {\doibase
  10.1002/ange.201604369} {\bibfield  {journal} {\bibinfo  {journal} {Angew.
  Chem.}\ }\textbf {\bibinfo {volume} {128}},\ \bibinfo {pages} {10448}
  (\bibinfo {year} {2016})}\BibitemShut {NoStop}%
\bibitem [{\citenamefont {Xu}\ \emph {et~al.}(2016)\citenamefont {Xu},
  \citenamefont {Du},\ and\ \citenamefont {Kou}}]{b2h2xu}%
  \BibitemOpen
  \bibfield  {author} {\bibinfo {author} {\bibfnamefont {L.-C.}\ \bibnamefont
  {Xu}}, \bibinfo {author} {\bibfnamefont {A.}~\bibnamefont {Du}}, \ and\
  \bibinfo {author} {\bibfnamefont {L.}~\bibnamefont {Kou}},\ }\href
  {http://pubs.rsc.org/en/content/articlelanding/2016/cp/c6cp05405f/unauth#!divAbstract}
  {\bibfield  {journal} {\bibinfo  {journal} {Phys. Chem. Chem. Phys.}\
  }\textbf {\bibinfo {volume} {18}},\ \bibinfo {pages} {27284} (\bibinfo {year}
  {2016})}\BibitemShut {NoStop}%
\bibitem [{\citenamefont {Nakhaee}\ \emph {et~al.}(2018)\citenamefont
  {Nakhaee}, \citenamefont {Ketabi},\ and\ \citenamefont
  {Peeters}}]{prb2018B2H2}%
  \BibitemOpen
  \bibfield  {author} {\bibinfo {author} {\bibfnamefont {M.}~\bibnamefont
  {Nakhaee}}, \bibinfo {author} {\bibfnamefont {S.A.}~\bibnamefont {Ketabi}}, \
  and\ \bibinfo {author} {\bibfnamefont {F.M.}~\bibnamefont {Peeters}},\ }\href
  {https://journals.aps.org/prb/pdf/10.1103/PhysRevB.97.125424} {\bibfield
  {journal} {\bibinfo  {journal} {Phys. Rev. B}\ }\textbf {\bibinfo {volume}
  {97}},\ \bibinfo {pages} {125424} (\bibinfo {year} {2018})}\BibitemShut
  {NoStop}%
\bibitem [{\citenamefont {Fu}\ and\ \citenamefont
  {Kane}(2007)}]{FuLparity2007}%
  \BibitemOpen
  \bibfield  {author} {\bibinfo {author} {\bibfnamefont {L.}~\bibnamefont
  {Fu}}\ and\ \bibinfo {author} {\bibfnamefont {C.~L.}\ \bibnamefont {Kane}},\
  }\href@noop {} {\bibfield  {journal} {\bibinfo  {journal} {Phys. Rev. B}\
  }\textbf {\bibinfo {volume} {76}},\ \bibinfo {pages} {045302} (\bibinfo
  {year} {2007})}\BibitemShut {NoStop}%
\bibitem [{\citenamefont {Mannix}\ \emph {et~al.}(2015)\citenamefont {Mannix},
  \citenamefont {Zhou}, \citenamefont {Kiraly}, \citenamefont {Wood},
  \citenamefont {Alducin}, \citenamefont {Myers}, \citenamefont {Liu},
  \citenamefont {Fisher}, \citenamefont {Santiago}, \citenamefont {Guest},
  \citenamefont {Yacaman}, \citenamefont {Ponce}, \citenamefont {Oganov},
  \citenamefont {Hersam},\ and\ \citenamefont {Guisinger}}]{borophenesy}%
  \BibitemOpen
  \bibfield  {author} {\bibinfo {author} {\bibfnamefont {A.~J.}\ \bibnamefont
  {Mannix}}, \bibinfo {author} {\bibfnamefont {X.-F.}\ \bibnamefont {Zhou}},
  \bibinfo {author} {\bibfnamefont {B.}~\bibnamefont {Kiraly}}, \bibinfo
  {author} {\bibfnamefont {J.~D.}\ \bibnamefont {Wood}}, \bibinfo {author}
  {\bibfnamefont {D.}~\bibnamefont {Alducin}}, \bibinfo {author} {\bibfnamefont
  {B.~D.}\ \bibnamefont {Myers}}, \bibinfo {author} {\bibfnamefont
  {X.}~\bibnamefont {Liu}}, \bibinfo {author} {\bibfnamefont {B.~L.}\
  \bibnamefont {Fisher}}, \bibinfo {author} {\bibfnamefont {U.}~\bibnamefont
  {Santiago}}, \bibinfo {author} {\bibfnamefont {J.~R.}\ \bibnamefont {Guest}},
  \bibinfo {author} {\bibfnamefont {M.~J.}\ \bibnamefont {Yacaman}}, \bibinfo
  {author} {\bibfnamefont {A.}~\bibnamefont {Ponce}}, \bibinfo {author}
  {\bibfnamefont {A.~R.}\ \bibnamefont {Oganov}}, \bibinfo {author}
  {\bibfnamefont {M.~C.}\ \bibnamefont {Hersam}}, \ and\ \bibinfo {author}
  {\bibfnamefont {N.~P.}\ \bibnamefont {Guisinger}},\ }\href {\doibase
  10.1126/science.aad1080} {\bibfield  {journal} {\bibinfo  {journal}
  {Science}\ }\textbf {\bibinfo {volume} {350}},\ \bibinfo {pages} {1513}
  (\bibinfo {year} {2015})}\BibitemShut {NoStop}%
\bibitem [{\citenamefont {Feng}\ \emph {et~al.}(2016)\citenamefont {Feng},
  \citenamefont {Zhang}, \citenamefont {Zhong}, \citenamefont {Li},
  \citenamefont {Li}, \citenamefont {Li}, \citenamefont {Cheng}, \citenamefont
  {Meng}, \citenamefont {Chen},\ and\ \citenamefont {Wu}}]{fengsy}%
  \BibitemOpen
  \bibfield  {author} {\bibinfo {author} {\bibfnamefont {B.}~\bibnamefont
  {Feng}}, \bibinfo {author} {\bibfnamefont {J.}~\bibnamefont {Zhang}},
  \bibinfo {author} {\bibfnamefont {Q.}~\bibnamefont {Zhong}}, \bibinfo
  {author} {\bibfnamefont {W.}~\bibnamefont {Li}}, \bibinfo {author}
  {\bibfnamefont {S.}~\bibnamefont {Li}}, \bibinfo {author} {\bibfnamefont
  {H.}~\bibnamefont {Li}}, \bibinfo {author} {\bibfnamefont {P.}~\bibnamefont
  {Cheng}}, \bibinfo {author} {\bibfnamefont {S.}~\bibnamefont {Meng}},
  \bibinfo {author} {\bibfnamefont {L.}~\bibnamefont {Chen}}, \ and\ \bibinfo
  {author} {\bibfnamefont {K.}~\bibnamefont {Wu}},\ }\href
  {https://www.nature.com/articles/nchem.2491} {\bibfield  {journal} {\bibinfo
  {journal} {Nat. Chem.}\ }\textbf {\bibinfo {volume} {8}},\ \bibinfo {pages}
  {563} (\bibinfo {year} {2016})}\BibitemShut {NoStop}%
\bibitem [{\citenamefont {Zhang}\ \emph
  {et~al.}(2016{\natexlab{a}})\citenamefont {Zhang}, \citenamefont {Penev},\
  and\ \citenamefont {Yakobson}}]{40}%
  \BibitemOpen
  \bibfield  {author} {\bibinfo {author} {\bibfnamefont {Z.}~\bibnamefont
  {Zhang}}, \bibinfo {author} {\bibfnamefont {E.~S.}\ \bibnamefont {Penev}}, \
  and\ \bibinfo {author} {\bibfnamefont {B.~I.}\ \bibnamefont {Yakobson}},\
  }\href {https://www.nature.com/articles/nchem.2521} {\bibfield  {journal}
  {\bibinfo  {journal} {Nat. Chem.}\ }\textbf {\bibinfo {volume} {8}},\
  \bibinfo {pages} {525} (\bibinfo {year} {2016}{\natexlab{a}})}\BibitemShut
  {NoStop}%
\bibitem [{\citenamefont {Zhang}\ \emph
  {et~al.}(2016{\natexlab{b}})\citenamefont {Zhang}, \citenamefont {Mannix},
  \citenamefont {Hu}, \citenamefont {Kiraly}, \citenamefont {Guisinger},
  \citenamefont {Hersam},\ and\ \citenamefont {Yakobson}}]{zhang2016sy}%
  \BibitemOpen
  \bibfield  {author} {\bibinfo {author} {\bibfnamefont {Z.}~\bibnamefont
  {Zhang}}, \bibinfo {author} {\bibfnamefont {A.~J.}\ \bibnamefont {Mannix}},
  \bibinfo {author} {\bibfnamefont {Z.}~\bibnamefont {Hu}}, \bibinfo {author}
  {\bibfnamefont {B.}~\bibnamefont {Kiraly}}, \bibinfo {author} {\bibfnamefont
  {N.~P.}\ \bibnamefont {Guisinger}}, \bibinfo {author} {\bibfnamefont {M.~C.}\
  \bibnamefont {Hersam}}, \ and\ \bibinfo {author} {\bibfnamefont {B.~I.}\
  \bibnamefont {Yakobson}},\ }\href
  {https://pubs.acs.org/doi/abs/10.1021/acs.nanolett.6b03349} {\bibfield
  {journal} {\bibinfo  {journal} {Nano Lett.}\ }\textbf {\bibinfo {volume}
  {16}},\ \bibinfo {pages} {6622} (\bibinfo {year}
  {2016}{\natexlab{b}})}\BibitemShut {NoStop}%
\bibitem [{\citenamefont {Zhong}\ \emph
  {et~al.}(2017{\natexlab{a}})\citenamefont {Zhong}, \citenamefont {Zhang},
  \citenamefont {Cheng}, \citenamefont {Feng}, \citenamefont {Li},
  \citenamefont {Sheng}, \citenamefont {Li}, \citenamefont {Meng},
  \citenamefont {Chen},\ and\ \citenamefont {Wu}}]{zhong2017sy}%
  \BibitemOpen
  \bibfield  {author} {\bibinfo {author} {\bibfnamefont {Q.}~\bibnamefont
  {Zhong}}, \bibinfo {author} {\bibfnamefont {J.}~\bibnamefont {Zhang}},
  \bibinfo {author} {\bibfnamefont {P.}~\bibnamefont {Cheng}}, \bibinfo
  {author} {\bibfnamefont {B.}~\bibnamefont {Feng}}, \bibinfo {author}
  {\bibfnamefont {W.}~\bibnamefont {Li}}, \bibinfo {author} {\bibfnamefont
  {S.}~\bibnamefont {Sheng}}, \bibinfo {author} {\bibfnamefont
  {H.}~\bibnamefont {Li}}, \bibinfo {author} {\bibfnamefont {S.}~\bibnamefont
  {Meng}}, \bibinfo {author} {\bibfnamefont {L.}~\bibnamefont {Chen}}, \ and\
  \bibinfo {author} {\bibfnamefont {K.}~\bibnamefont {Wu}},\ }\href@noop {}
  {\bibfield  {journal} {\bibinfo  {journal} {J. Phy-Condens. Mat.}\ }\textbf
  {\bibinfo {volume} {29}},\ \bibinfo {pages} {095002} (\bibinfo {year}
  {2017}{\natexlab{a}})}\BibitemShut {NoStop}%
\bibitem [{\citenamefont {Zhong}\ \emph
  {et~al.}(2017{\natexlab{b}})\citenamefont {Zhong}, \citenamefont {Kong},
  \citenamefont {Gou}, \citenamefont {Li}, \citenamefont {Sheng}, \citenamefont
  {Yang}, \citenamefont {Cheng}, \citenamefont {Li}, \citenamefont {Wu},\ and\
  \citenamefont {Chen}}]{prmribbbionsy}%
  \BibitemOpen
  \bibfield  {author} {\bibinfo {author} {\bibfnamefont {Q.}~\bibnamefont
  {Zhong}}, \bibinfo {author} {\bibfnamefont {L.}~\bibnamefont {Kong}},
  \bibinfo {author} {\bibfnamefont {J.}~\bibnamefont {Gou}}, \bibinfo {author}
  {\bibfnamefont {W.}~\bibnamefont {Li}}, \bibinfo {author} {\bibfnamefont
  {S.}~\bibnamefont {Sheng}}, \bibinfo {author} {\bibfnamefont
  {S.}~\bibnamefont {Yang}}, \bibinfo {author} {\bibfnamefont {P.}~\bibnamefont
  {Cheng}}, \bibinfo {author} {\bibfnamefont {H.}~\bibnamefont {Li}}, \bibinfo
  {author} {\bibfnamefont {K.}~\bibnamefont {Wu}}, \ and\ \bibinfo {author}
  {\bibfnamefont {L.}~\bibnamefont {Chen}},\ }\href {\doibase
  10.1103/PhysRevMaterials.1.021001} {\bibfield  {journal} {\bibinfo  {journal}
  {Phys. Rev. Mater.}\ }\textbf {\bibinfo {volume} {1}},\ \bibinfo {pages}
  {021001} (\bibinfo {year} {2017}{\natexlab{b}})}\BibitemShut {NoStop}%
\bibitem [{\citenamefont {Li}\ \emph {et~al.}(2018)\citenamefont {Li},
  \citenamefont {Kong}, \citenamefont {Chen}, \citenamefont {Gou},
  \citenamefont {Sheng}, \citenamefont {Zhang}, \citenamefont {Li},
  \citenamefont {Chen}, \citenamefont {Cheng},\ and\ \citenamefont
  {Wu}}]{wukehui}%
  \BibitemOpen
  \bibfield  {author} {\bibinfo {author} {\bibfnamefont {W.}~\bibnamefont
  {Li}}, \bibinfo {author} {\bibfnamefont {L.}~\bibnamefont {Kong}}, \bibinfo
  {author} {\bibfnamefont {C.}~\bibnamefont {Chen}}, \bibinfo {author}
  {\bibfnamefont {J.}~\bibnamefont {Gou}}, \bibinfo {author} {\bibfnamefont
  {S.}~\bibnamefont {Sheng}}, \bibinfo {author} {\bibfnamefont
  {W.}~\bibnamefont {Zhang}}, \bibinfo {author} {\bibfnamefont
  {H.}~\bibnamefont {Li}}, \bibinfo {author} {\bibfnamefont {L.}~\bibnamefont
  {Chen}}, \bibinfo {author} {\bibfnamefont {P.}~\bibnamefont {Cheng}}, \ and\
  \bibinfo {author} {\bibfnamefont {K.}~\bibnamefont {Wu}},\ }\href {\doibase
  https://doi.org/10.1016/j.scib.2018.02.006} {\bibfield  {journal} {\bibinfo
  {journal} {Sci. Bull.}\ }\textbf {\bibinfo {volume} {63}},\ \bibinfo {pages}
  {282 } (\bibinfo {year} {2018})}\BibitemShut {NoStop}%
\bibitem [{\citenamefont {Feng}\ \emph
  {et~al.}(2017{\natexlab{b}})\citenamefont {Feng}, \citenamefont {Sugino},
  \citenamefont {Liu}, \citenamefont {Zhang}, \citenamefont {Yukawa},
  \citenamefont {Kawamura}, \citenamefont {Iimori}, \citenamefont {Kim},
  \citenamefont {Hasegawa}, \citenamefont {Li}, \citenamefont {Chen},
  \citenamefont {Wu}, \citenamefont {Kumigashira}, \citenamefont {Komori},
  \citenamefont {Chiang}, \citenamefont {Meng},\ and\ \citenamefont
  {Matsuda}}]{fengprl}%
  \BibitemOpen
  \bibfield  {author} {\bibinfo {author} {\bibfnamefont {B.}~\bibnamefont
  {Feng}}, \bibinfo {author} {\bibfnamefont {O.}~\bibnamefont {Sugino}},
  \bibinfo {author} {\bibfnamefont {R.-Y.}\ \bibnamefont {Liu}}, \bibinfo
  {author} {\bibfnamefont {J.}~\bibnamefont {Zhang}}, \bibinfo {author}
  {\bibfnamefont {R.}~\bibnamefont {Yukawa}}, \bibinfo {author} {\bibfnamefont
  {M.}~\bibnamefont {Kawamura}}, \bibinfo {author} {\bibfnamefont
  {T.}~\bibnamefont {Iimori}}, \bibinfo {author} {\bibfnamefont
  {H.}~\bibnamefont {Kim}}, \bibinfo {author} {\bibfnamefont {Y.}~\bibnamefont
  {Hasegawa}}, \bibinfo {author} {\bibfnamefont {H.}~\bibnamefont {Li}},
  \bibinfo {author} {\bibfnamefont {L.}~\bibnamefont {Chen}}, \bibinfo {author}
  {\bibfnamefont {K.}~\bibnamefont {Wu}}, \bibinfo {author} {\bibfnamefont
  {H.}~\bibnamefont {Kumigashira}}, \bibinfo {author} {\bibfnamefont
  {F.}~\bibnamefont {Komori}}, \bibinfo {author} {\bibfnamefont {T.-C.}\
  \bibnamefont {Chiang}}, \bibinfo {author} {\bibfnamefont {S.}~\bibnamefont
  {Meng}}, \ and\ \bibinfo {author} {\bibfnamefont {I.}~\bibnamefont
  {Matsuda}},\ }\href {\doibase 10.1103/PhysRevLett.118.096401} {\bibfield
  {journal} {\bibinfo  {journal} {Phys. Rev. Lett.}\ }\textbf {\bibinfo
  {volume} {118}},\ \bibinfo {pages} {096401} (\bibinfo {year}
  {2017}{\natexlab{b}})}\BibitemShut {NoStop}%
\bibitem [{\citenamefont {Feng}\ \emph {et~al.}(2018)\citenamefont {Feng},
  \citenamefont {Zhang}, \citenamefont {Ito}, \citenamefont {Arita},
  \citenamefont {Cheng}, \citenamefont {Chen}, \citenamefont {Wu},
  \citenamefont {Komori}, \citenamefont {Sugino}, \citenamefont {Miyamoto}
  \emph {et~al.}}]{feng2018discovery}%
  \BibitemOpen
  \bibfield  {author} {\bibinfo {author} {\bibfnamefont {B.}~\bibnamefont
  {Feng}}, \bibinfo {author} {\bibfnamefont {J.}~\bibnamefont {Zhang}},
  \bibinfo {author} {\bibfnamefont {S.}~\bibnamefont {Ito}}, \bibinfo {author}
  {\bibfnamefont {M.}~\bibnamefont {Arita}}, \bibinfo {author} {\bibfnamefont
  {C.}~\bibnamefont {Cheng}}, \bibinfo {author} {\bibfnamefont
  {L.}~\bibnamefont {Chen}}, \bibinfo {author} {\bibfnamefont {K.}~\bibnamefont
  {Wu}}, \bibinfo {author} {\bibfnamefont {F.}~\bibnamefont {Komori}}, \bibinfo
  {author} {\bibfnamefont {O.}~\bibnamefont {Sugino}}, \bibinfo {author}
  {\bibfnamefont {K.}~\bibnamefont {Miyamoto}},  \emph {et~al.},\ }\href@noop
  {} {\bibfield  {journal} {\bibinfo  {journal} {Adv. Mater.}\ }\textbf
  {\bibinfo {volume} {30}},\ \bibinfo {pages} {1704025} (\bibinfo {year}
  {2018})}\BibitemShut {NoStop}%
\bibitem [{\citenamefont {Martinez-Canales}\ \emph {et~al.}(2017)\citenamefont
  {Martinez-Canales}, \citenamefont {Galeev}, \citenamefont {Boldyrev},\ and\
  \citenamefont {Pickard}}]{prbborane}%
  \BibitemOpen
  \bibfield  {author} {\bibinfo {author} {\bibfnamefont {M.}~\bibnamefont
  {Martinez-Canales}}, \bibinfo {author} {\bibfnamefont {T.~R.}\ \bibnamefont
  {Galeev}}, \bibinfo {author} {\bibfnamefont {A.~I.}\ \bibnamefont
  {Boldyrev}}, \ and\ \bibinfo {author} {\bibfnamefont {C.~J.}\ \bibnamefont
  {Pickard}},\ }\href {\doibase 10.1103/PhysRevB.96.195442} {\bibfield
  {journal} {\bibinfo  {journal} {Phys. Rev. B}\ }\textbf {\bibinfo {volume}
  {96}},\ \bibinfo {pages} {195442} (\bibinfo {year} {2017})}\BibitemShut
  {NoStop}%
\bibitem [{\citenamefont {Ma}\ \emph {et~al.}(2016)\citenamefont {Ma},
  \citenamefont {Jiao}, \citenamefont {Gao}, \citenamefont {Gu}, \citenamefont
  {Bilic}, \citenamefont {Chen},\ and\ \citenamefont {Du}}]{ma2016graphene}%
  \BibitemOpen
  \bibfield  {author} {\bibinfo {author} {\bibfnamefont {F.}~\bibnamefont
  {Ma}}, \bibinfo {author} {\bibfnamefont {Y.}~\bibnamefont {Jiao}}, \bibinfo
  {author} {\bibfnamefont {G.}~\bibnamefont {Gao}}, \bibinfo {author}
  {\bibfnamefont {Y.}~\bibnamefont {Gu}}, \bibinfo {author} {\bibfnamefont
  {A.}~\bibnamefont {Bilic}}, \bibinfo {author} {\bibfnamefont
  {Z.}~\bibnamefont {Chen}}, \ and\ \bibinfo {author} {\bibfnamefont
  {A.}~\bibnamefont {Du}},\ }\href@noop {} {\bibfield  {journal} {\bibinfo
  {journal} {Nano Lett.}\ }\textbf {\bibinfo {volume} {16}},\ \bibinfo {pages}
  {3022} (\bibinfo {year} {2016})}\BibitemShut {NoStop}%
\bibitem [{\citenamefont {Zhang}\ \emph
  {et~al.}(2016{\natexlab{c}})\citenamefont {Zhang}, \citenamefont {Li},
  \citenamefont {Hou}, \citenamefont {Du},\ and\ \citenamefont
  {Chen}}]{zhang2016dirac}%
  \BibitemOpen
  \bibfield  {author} {\bibinfo {author} {\bibfnamefont {H.}~\bibnamefont
  {Zhang}}, \bibinfo {author} {\bibfnamefont {Y.}~\bibnamefont {Li}}, \bibinfo
  {author} {\bibfnamefont {J.}~\bibnamefont {Hou}}, \bibinfo {author}
  {\bibfnamefont {A.}~\bibnamefont {Du}}, \ and\ \bibinfo {author}
  {\bibfnamefont {Z.}~\bibnamefont {Chen}},\ }\href
  {https://pubs.acs.org/doi/abs/10.1021/acs.nanolett.6b02335} {\bibfield
  {journal} {\bibinfo  {journal} {Nano Lett.}\ }\textbf {\bibinfo {volume}
  {16}},\ \bibinfo {pages} {6124} (\bibinfo {year}
  {2016}{\natexlab{c}})}\BibitemShut {NoStop}%
\bibitem [{\citenamefont {Zhang}\ \emph {et~al.}(2014)\citenamefont {Zhang},
  \citenamefont {Wang}, \citenamefont {Du}, \citenamefont {Gao},\ and\
  \citenamefont {Liu}}]{TiB2dirac}%
  \BibitemOpen
  \bibfield  {author} {\bibinfo {author} {\bibfnamefont {L.~Z.}\ \bibnamefont
  {Zhang}}, \bibinfo {author} {\bibfnamefont {Z.~F.}\ \bibnamefont {Wang}},
  \bibinfo {author} {\bibfnamefont {S.~X.}\ \bibnamefont {Du}}, \bibinfo
  {author} {\bibfnamefont {H.-J.}\ \bibnamefont {Gao}}, \ and\ \bibinfo
  {author} {\bibfnamefont {F.}~\bibnamefont {Liu}},\ }\href {\doibase
  10.1103/PhysRevB.90.161402} {\bibfield  {journal} {\bibinfo  {journal} {Phys.
  Rev. B}\ }\textbf {\bibinfo {volume} {90}},\ \bibinfo {pages} {161402}
  (\bibinfo {year} {2014})}\BibitemShut {NoStop}%
\bibitem [{\citenamefont {Kresse}\ and\ \citenamefont {Hafner}(1993)}]{28}%
  \BibitemOpen
  \bibfield  {author} {\bibinfo {author} {\bibfnamefont {G.}~\bibnamefont
  {Kresse}}\ and\ \bibinfo {author} {\bibfnamefont {J.}~\bibnamefont
  {Hafner}},\ }\href {\doibase 10.1103/PhysRevB.47.558} {\bibfield  {journal}
  {\bibinfo  {journal} {Phys. Rev. B}\ }\textbf {\bibinfo {volume} {47}},\
  \bibinfo {pages} {558} (\bibinfo {year} {1993})}\BibitemShut {NoStop}%
\bibitem [{\citenamefont {Bl\"ochl}(1994)}]{32}%
  \BibitemOpen
  \bibfield  {author} {\bibinfo {author} {\bibfnamefont {P.~E.}\ \bibnamefont
  {Bl\"ochl}},\ }\href {\doibase 10.1103/PhysRevB.50.17953} {\bibfield
  {journal} {\bibinfo  {journal} {Phys. Rev. B}\ }\textbf {\bibinfo {volume}
  {50}},\ \bibinfo {pages} {17953} (\bibinfo {year} {1994})}\BibitemShut
  {NoStop}%
\bibitem [{\citenamefont {Perdew}\ \emph {et~al.}(1992)\citenamefont {Perdew},
  \citenamefont {Chevary}, \citenamefont {Vosko}, \citenamefont {Jackson},
  \citenamefont {Pederson}, \citenamefont {Singh},\ and\ \citenamefont
  {Fiolhais}}]{33}%
  \BibitemOpen
  \bibfield  {author} {\bibinfo {author} {\bibfnamefont {J.~P.}\ \bibnamefont
  {Perdew}}, \bibinfo {author} {\bibfnamefont {J.~A.}\ \bibnamefont {Chevary}},
  \bibinfo {author} {\bibfnamefont {S.~H.}\ \bibnamefont {Vosko}}, \bibinfo
  {author} {\bibfnamefont {K.~A.}\ \bibnamefont {Jackson}}, \bibinfo {author}
  {\bibfnamefont {M.~R.}\ \bibnamefont {Pederson}}, \bibinfo {author}
  {\bibfnamefont {D.~J.}\ \bibnamefont {Singh}}, \ and\ \bibinfo {author}
  {\bibfnamefont {C.}~\bibnamefont {Fiolhais}},\ }\href {\doibase
  10.1103/PhysRevB.46.6671} {\bibfield  {journal} {\bibinfo  {journal} {Phys.
  Rev. B}\ }\textbf {\bibinfo {volume} {46}},\ \bibinfo {pages} {6671}
  (\bibinfo {year} {1992})}\BibitemShut {NoStop}%
\bibitem [{\citenamefont {Perdew}\ \emph {et~al.}(1997)\citenamefont {Perdew},
  \citenamefont {Burke},\ and\ \citenamefont {Ernzerhof}}]{29}%
  \BibitemOpen
  \bibfield  {author} {\bibinfo {author} {\bibfnamefont {J.~P.}\ \bibnamefont
  {Perdew}}, \bibinfo {author} {\bibfnamefont {K.}~\bibnamefont {Burke}}, \
  and\ \bibinfo {author} {\bibfnamefont {M.}~\bibnamefont {Ernzerhof}},\ }\href
  {\doibase 10.1103/PhysRevLett.78.1396} {\bibfield  {journal} {\bibinfo
  {journal} {Phys. Rev. Lett.}\ }\textbf {\bibinfo {volume} {78}},\ \bibinfo
  {pages} {1396} (\bibinfo {year} {1997})}\BibitemShut {NoStop}%
\bibitem [{\citenamefont {Marzari}\ and\ \citenamefont
  {Vanderbilt}(1997)}]{30}%
  \BibitemOpen
  \bibfield  {author} {\bibinfo {author} {\bibfnamefont {N.}~\bibnamefont
  {Marzari}}\ and\ \bibinfo {author} {\bibfnamefont {D.}~\bibnamefont
  {Vanderbilt}},\ }\href {\doibase 10.1103/PhysRevB.56.12847} {\bibfield
  {journal} {\bibinfo  {journal} {Phys. Rev. B}\ }\textbf {\bibinfo {volume}
  {56}},\ \bibinfo {pages} {12847} (\bibinfo {year} {1997})}\BibitemShut
  {NoStop}%
\bibitem [{\citenamefont {Souza}\ \emph {et~al.}(2001)\citenamefont {Souza},
  \citenamefont {Marzari},\ and\ \citenamefont {Vanderbilt}}]{31}%
  \BibitemOpen
  \bibfield  {author} {\bibinfo {author} {\bibfnamefont {I.}~\bibnamefont
  {Souza}}, \bibinfo {author} {\bibfnamefont {N.}~\bibnamefont {Marzari}}, \
  and\ \bibinfo {author} {\bibfnamefont {D.}~\bibnamefont {Vanderbilt}},\
  }\href {\doibase 10.1103/PhysRevB.65.035109} {\bibfield  {journal} {\bibinfo
  {journal} {Phys. Rev. B}\ }\textbf {\bibinfo {volume} {65}},\ \bibinfo
  {pages} {035109} (\bibinfo {year} {2001})}\BibitemShut {NoStop}%
\bibitem [{\citenamefont {Kariyado}\ and\ \citenamefont {Hatsugai}(2013)}]{35}%
  \BibitemOpen
  \bibfield  {author} {\bibinfo {author} {\bibfnamefont {T.}~\bibnamefont
  {Kariyado}}\ and\ \bibinfo {author} {\bibfnamefont {Y.}~\bibnamefont
  {Hatsugai}},\ }\href {\doibase 10.1103/PhysRevB.88.245126} {\bibfield
  {journal} {\bibinfo  {journal} {Phys. Rev. B}\ }\textbf {\bibinfo {volume}
  {88}},\ \bibinfo {pages} {245126} (\bibinfo {year} {2013})}\BibitemShut
  {NoStop}%
\bibitem [{\citenamefont {Fu}\ and\ \citenamefont {Kane}(2006)}]{36}%
  \BibitemOpen
  \bibfield  {author} {\bibinfo {author} {\bibfnamefont {L.}~\bibnamefont
  {Fu}}\ and\ \bibinfo {author} {\bibfnamefont {C.~L.}\ \bibnamefont {Kane}},\
  }\href {\doibase 10.1103/PhysRevB.74.195312} {\bibfield  {journal} {\bibinfo
  {journal} {Phys. Rev. B}\ }\textbf {\bibinfo {volume} {74}},\ \bibinfo
  {pages} {195312} (\bibinfo {year} {2006})}\BibitemShut {NoStop}%
\bibitem [{\citenamefont {Kim}\ \emph {et~al.}(2015{\natexlab{b}})\citenamefont
  {Kim}, \citenamefont {Baik}, \citenamefont {Ryu}, \citenamefont {Sohn},
  \citenamefont {Park}, \citenamefont {Park}, \citenamefont {Denlinger},
  \citenamefont {Yi}, \citenamefont {Choi},\ and\ \citenamefont
  {Kim}}]{kim2015observation}%
  \BibitemOpen
  \bibfield  {author} {\bibinfo {author} {\bibfnamefont {J.}~\bibnamefont
  {Kim}}, \bibinfo {author} {\bibfnamefont {S.~S.}\ \bibnamefont {Baik}},
  \bibinfo {author} {\bibfnamefont {S.~H.}\ \bibnamefont {Ryu}}, \bibinfo
  {author} {\bibfnamefont {Y.}~\bibnamefont {Sohn}}, \bibinfo {author}
  {\bibfnamefont {S.}~\bibnamefont {Park}}, \bibinfo {author} {\bibfnamefont
  {B.-G.}\ \bibnamefont {Park}}, \bibinfo {author} {\bibfnamefont
  {J.}~\bibnamefont {Denlinger}}, \bibinfo {author} {\bibfnamefont
  {Y.}~\bibnamefont {Yi}}, \bibinfo {author} {\bibfnamefont {H.~J.}\
  \bibnamefont {Choi}}, \ and\ \bibinfo {author} {\bibfnamefont {K.~S.}\
  \bibnamefont {Kim}},\ }\href {\doibase 10.1126/science.aaa6486} {\bibfield
  {journal} {\bibinfo  {journal} {Science}\ }\textbf {\bibinfo {volume}
  {349}},\ \bibinfo {pages} {723} (\bibinfo {year}
  {2015}{\natexlab{b}})}\BibitemShut {NoStop}%
\bibitem [{\citenamefont {Qian}\ \emph {et~al.}(2014)\citenamefont {Qian},
  \citenamefont {Liu}, \citenamefont {Fu},\ and\ \citenamefont
  {Li}}]{qian2014quantum}%
  \BibitemOpen
  \bibfield  {author} {\bibinfo {author} {\bibfnamefont {X.}~\bibnamefont
  {Qian}}, \bibinfo {author} {\bibfnamefont {J.}~\bibnamefont {Liu}}, \bibinfo
  {author} {\bibfnamefont {L.}~\bibnamefont {Fu}}, \ and\ \bibinfo {author}
  {\bibfnamefont {J.}~\bibnamefont {Li}},\ }\href {\doibase
  10.1126/science.1256815} {\bibfield  {journal} {\bibinfo  {journal}
  {Science}\ }\textbf {\bibinfo {volume} {346}},\ \bibinfo {pages} {1344}
  (\bibinfo {year} {2014})}\BibitemShut {NoStop}%
\bibitem [{\citenamefont {Wu}\ \emph {et~al.}(2018)\citenamefont {Wu},
  \citenamefont {Fatemi}, \citenamefont {Gibson}, \citenamefont {Watanabe},
  \citenamefont {Taniguchi}, \citenamefont {Cava},\ and\ \citenamefont
  {Jarillo-Herrero}}]{wu2018observation}%
  \BibitemOpen
  \bibfield  {author} {\bibinfo {author} {\bibfnamefont {S.}~\bibnamefont
  {Wu}}, \bibinfo {author} {\bibfnamefont {V.}~\bibnamefont {Fatemi}}, \bibinfo
  {author} {\bibfnamefont {Q.~D.}\ \bibnamefont {Gibson}}, \bibinfo {author}
  {\bibfnamefont {K.}~\bibnamefont {Watanabe}}, \bibinfo {author}
  {\bibfnamefont {T.}~\bibnamefont {Taniguchi}}, \bibinfo {author}
  {\bibfnamefont {R.~J.}\ \bibnamefont {Cava}}, \ and\ \bibinfo {author}
  {\bibfnamefont {P.}~\bibnamefont {Jarillo-Herrero}},\ }\href {\doibase
  10.1126/science.aan6003} {\bibfield  {journal} {\bibinfo  {journal}
  {Science}\ }\textbf {\bibinfo {volume} {359}},\ \bibinfo {pages} {76}
  (\bibinfo {year} {2018})}\BibitemShut {NoStop}%
\end{thebibliography}%

\end{document}